\documentclass[english,twocolumn, superscriptaddress, nofootinbib]{revtex4-1}
\usepackage{ae,aecompl}
\usepackage[T1]{fontenc}
\usepackage[latin9]{inputenc}
\usepackage[pdftex]{geometry}
\usepackage[pdftex]{xcolor}
\usepackage{pdfcolmk}
\usepackage{babel}
\usepackage{float}
\usepackage{units}
\usepackage{textcomp}
\usepackage{mathtools}
\usepackage{bm}
\usepackage{amsmath}
\usepackage{amssymb}
\usepackage[pdftex]{graphicx}
\usepackage{wasysym}
\PassOptionsToPackage{normalem}{ulem}
\usepackage{ulem}
\usepackage[unicode=true,pdfusetitle,
 bookmarks=true,bookmarksnumbered=false,bookmarksopen=false,
 breaklinks=false,pdfborder={0 0 0},pdfborderstyle={},backref=false,colorlinks=false]
 {hyperref}
\hypersetup{
 colorlinks = true, citecolor = blue, urlcolor = cyan, linkcolor = magenta}

\makeatletter

\providecolor{lyxadded}{rgb}{0,0,1}
\providecolor{lyxdeleted}{rgb}{1,0,0}
\DeclareRobustCommand{\lyxsout}[1]{\ifx\\#1\else\sout{#1}\fi}

\usepackage[bottom]{footmisc}

\makeatother

\begin{document}
\global\long\def\vect#1{\overrightarrow{\mathbf{#1}}}%

\global\long\def\abs#1{\left|#1\right|}%

\global\long\def\av#1{\left\langle #1\right\rangle }%

\global\long\def\ket#1{\left|#1\right\rangle }%

\global\long\def\bra#1{\left\langle #1\right|}%

\global\long\def\tensorproduct{\otimes}%

\global\long\def\braket#1#2{\left\langle #1\mid#2\right\rangle }%

\global\long\def\omv{\overrightarrow{\Omega}}%

\global\long\def\inf{\infty}%

\title{Landauer transport as a quasisteady state on finite chains under unitary
quantum dynamics}
\author{J. P. Santos Pires}
\email{up201201453@fc.up.pt}

\affiliation{Centro de Física das Universidades do Minho e Porto~\\
Departamento de Física e Astronomia, Faculdade de Ciências, Universidade
do Porto, 4169-007 Porto, Portugal}
\author{B. Amorim}
\email{amorim.bac@gmail.com}

\affiliation{Centro de Física das Universidades do Minho e Porto~\\
University of Minho, Campus of Gualtar, 4710-057, Braga, Portugal}
\author{J. M. Viana Parente Lopes}
\email{jlopes@fc.up.pt}

\affiliation{Centro de Física das Universidades do Minho e Porto~\\
Departamento de Física e Astronomia, Faculdade de Ciências, Universidade
do Porto, 4169-007 Porto, Portugal}
\begin{abstract}
$ $

$\ \ \ \ \ \ \ \ \ \ \ \ \ \ \ \ \ \ \ \ \ \ \ \ \ \ \ \ \ $\textit{(Accepted
in the Physical Review B - March 2020)}

$ $

In this paper, we study the emergence of a Landauer transport regime
from the quantum-mechanical dynamics of free electrons in a disordered
tight-binding chain, which is coupled to finite leads with open boundaries.
Both\textit{\normalsize{} partitioned} and\textit{\normalsize{} partition-free}
initial conditions are analyzed and seen to give rise, for large enough
leads, to the same spatially uniform quasi-steady-state current, which
agrees with the Landauer value. The quasi-steady-state regime is preceded
by a transient regime, which lasts for a time proportional to the
length of the disordered sample, and followed by recursions, after
a time that is proportional to the lead size. These theoretical predictions
may be of interest to future experiments on transport of fermionic
ultra-cold atoms across optical lattices. We also observe finite-size
current oscillations, superimposed on the quasi-steady state, whose
behavior depends crucially on the conditions initially imposed on
the system. Finally, we show how a time-resolved Kubo formula is able
to reproduce this Landauer transport regime, as the leads grow bigger.

$ $
\end{abstract}
\maketitle

\section{\label{sec:Introduction}Introduction}

The study of electronic transport is amongst the main goals of condensed
matter physics. In the regime of small length scales and low temperatures,
the mesoscopic transport regime, quantum coherence effects play a
dominant role in the propagation of electron states. In such a case,
transport can no longer be seen as a bulk phenomenon, but instead
depends on device-specific details such as the geometry of and nature
of the electrodes, as well as the specific distribution of disorder
in the sample.

A theoretical description of mesoscopic transport was first developed
by Landauer \citep{landauer_electrical_1970} and later generalized
by Büttiker \citep{buttiker_four-terminal_1986}. In the now called
\textit{Landauer-Büttiker formalism}, the problem of stationary mesoscopic
transport is recast as a scattering problem, where single-electron
states incoming from the leads are transmitted across a central device.
The current may then be expressed as a sum over the transmission probabilities
of the occupied incoming lead states. In parallel to this work, Caroli
\citep{caroli_direct_1971} applied the non-equilibrium Green's function
formalism of Kadanoff-Baym \citep{Kadanoff_Baym_book} and Keldysh
\citep{Keldysh_1964} to the calculation of mesoscopic transport.
The obtained expression has a structure similar to the Landauer-Büttiker
(LB) one, but with the transmission coefficient now expressed in terms
of Green's functions of the central device and spectral functions
of the leads. While apparently distinct, the two approaches lead to
the same result, as implied by the Fisher-Lee relation \citep{fisher_relation_1981,stone_what_1988,baranger_electrical_1989}
between transmission coefficients and Green's functions (for a detailed
proof, see Wimmer \citep{wimmer_quantum_2009}). Central to both approaches
are the assumptions that the leads attached to the central device
are semi-infinite and the occupation of the incoming single-electron
states is determined by independent Fermi energies on each lead. Moreover,
both methods are only able to describe steady-state transport. We
also point out that the Landauer formula has also been derived within
the theory of non-equilibrium steady states \citep{nenciu_independent_2007}.

If one is interested in the transient dynamics and how this steady
state is reached, the matter of what the initial condition of the
system was, becomes relevant. At the theoretical level, two initial
conditions have been historically considered:\textit{ (i) }In the
\textit{partitioned approach} \citep{caroli_direct_1971,meir_landauer_1992,jauho_time-dependent_1994},
the leads and the central device are assumed to be initially disconnected,
each being in equilibrium with independent Fermi levels. This Fermi-level
imbalance takes into account the bias applied to the mesoscopic device.
Then, the leads and the device are suddenly brought into contact allowing
a charge current to flow. \textit{(ii)} In the \textit{partition-free
approach} \citep{Cini1980,stefanucci_time-dependent_2004,Odashima},
the leads and the device are assumed to be connected from the beginning
and in global equilibrium with a common Fermi energy. Then, a potential
bias between the leads is suddenly applied to the connected system.
It has been shown, for the case of a single-level central device,
that the same steady-state current is reached in both approaches,
for the same time-dependent perturbation \citep{stefanucci_time-dependent_2004}.
Furthermore, the value of the steady-state current does not depend
on the history of the time-dependent perturbation, provided it reaches
the same constant value in the future. Crucial to this result is the
fact that the leads have a continuum spectrum (as it occurs for semi-infinite
leads), which allows a loss of memory about the initial conditions,
provided there are no bound states in the central device. The existence
of bound states inside the device is known to cause persistent current
oscillations \citep{khosravi_role_2008,khosravi_bound_2009,cornean_memory_2014}.
We also point out that, under certain circumstances, interactions
might prevent the formation of a steady state \citep{khosravi_bound_2009}.
Finally, the equivalence of the steady-state current reached in the
\textit{partitioned} and \textit{partition-free} cases was shown,
with rather broad assumptions and including the presence of interactions,
in a mathematically rigorous way in Refs.\citep{moldoveanu_nonequilibrium_2011,cornean_adiabatic_2012,cornean_steady_2014}.

\textcolor{black}{Very recently, Purkayastha }\textit{\textcolor{black}{et
al}}\textcolor{black}{{} \citep{purkayastha_classifying_2019} also
discussed the problem of relating the linear response theory in a
general interacting system coupled infinite leads to the open system's
Kubo formula. The authors looked at the asymptotic behavior of the
integrated current-current correlators at different times, in the
limit of very large systems and times. This way, they found a surprising
disagreement between the transport classification given by both approaches
\citep{PhysRevB.97.174206} for the critical one-dimensional Aubry-André-Harper
model. This was attributed to the non-commutativity of the two referred
limits. Nevertheless, their analysis was always done in a }\textit{\textcolor{black}{partitioned}}\textcolor{black}{{}
setup and in linear response theory.}

In more recent years, a significant effort was devoted to the study
of time-dependent transport and transient dynamics in mesoscopic systems
attached to infinite leads \citep{jauho_time-dependent_1994,Stefanucci_2004,Tuovinen_2013,tuovinen_time-dependent_2014,Latini_2014,popescu_efficient_2016,popescu_emergence_2017}.
However, the investigation of time-dependent transport in systems
where the leads are finite (but possibly very large) has received
much less attention. Initial work on this problem was made in Refs.
\citep{ventra_transport_2004,bushong_approach_2005,chien_bosonic_2012},
in which a micro-canonical method was developed to deal with quasi-steady-state
transport in finite systems, where the leads are initially connected
to the device, but one of them is partially or fully depleted of particles.
More recently, the same problem was considered in Pal \textit{et al}
\citep{pal_emergence_2018}, which studied time-dependent transport
through a quantum dot connected to two systems with a quasi-continuum
spectrum (discrete, but dense), which take the role of finite leads,
in the \textit{partitioned }approach. In all of these works, it was
observed that after the transients died out, a quasisteady state transport
regime emerges. However, the dependence of the quasisteady state on
the system initial condition has never been explored so far. Is the
quasi-steady-state current independent of the initial preparation
of the system? Does the manner in which the quasisteady state is reached
depend on these initial conditions? These are the questions we try
to answer in this work. Notice that these issues are not of mere theoretical
interest. It is true that for realistic electronic devices, consideration
of finite leads is somewhat artificial.\footnote{Unless one is interested in times comparable to the discharge time
of a battery, the electrodes and voltage source that are connected
to a mesoscopic device can be seen as infinite.} However, cold atoms trapped in optical lattices have emerged as a
platform for the experimental study of transport properties \citep{ott_collisionally_2004,rom_free_2006}
{[}also studied theoretically in Chien\textit{ et al.} \citep{chien_bosonic_2012}{]}.
In these systems, due to the limited sizes of optical lattices, the
``leads'' are necessarily finite. Furthermore, while in realistic
low-dimensional electronic devices, one expects electron-electron
interactions play a significant role \citep{todorov_tight-binding_2002,bushong_approach_2005,KurthPRL2010,Latini_2014,cornean_steady_2014},
in optical lattices the inter-particle interaction can be tuned down
to zero, thus enabling a proper study of transport for non-interacting
fermions \citep{ott_collisionally_2004,rom_free_2006}. As such, we
expect that our theoretical results will be of experimental relevance
for transport in ultra-cold-atoms setups.

The purpose of this work is to further explore how a steady-state
transport regime emerges from quantum time-evolution in non-interacting,
fermionic systems with finite, but large leads, and how the initial
state of the system affects this process. By combining numerical and
analytical work, we study the time-dependent current dynamics in a
prototypical one-dimensional non-interacting tight-binding model with
disorder, analyzing in detail how the current dynamics depends on
the initial conditions (\textit{partitioned }vs\textit{ partition-free})
and on the size of the finite leads. We employ a full quantum time
evolution, starting from both initial conditions, to study the time-dependent
current upon the sudden connection of the appropriate perturbation.
For the\textit{ partition-free} case, we also derive a time-dependent
Kubo formula, which allows us to see rigorously how a linear Landauer-Büttiker
formula, involving only quantum transmittances, emerges from an unitary
time evolution in the limit of very large l\textcolor{black}{eads.}

The text is organized as follows. In Sec.~\ref{sec:Hamiltonian},
we introduce the one-dimensional tight-binding model Hamiltonian that
will be used throughout the rest of the paper and detail both the
\textit{partitioned} and \textit{partition-free} approaches. In Sec.~\ref{sec:Numerical-methods},
we describe the numerical methods used for calculating the time-dependent
local current from the unitary dynamics of the finite system and also
the steady-state Landauer current for infinite leads. The main numerical
results are then presented in Sec.~\ref{sec:NumericalResults}, where
the time-evolution of the non-equilibrium current is systematically
analyzed as a function of the bias, the size of the finite leads,
and the central sample's disorder and size. Finally, in Sec.~\ref{sec:Emergence-of-Landauer},
we provide analytical insight into the numerical results of Sec.~\ref{sec:NumericalResults},
by developing a time-dependent Kubo formula for the \textit{partition-free}
approach and expressing it in terms of complex reflection and transmission
coefficients of the central sample. In Sec.~\ref{sec:Conclusions},
we discuss the obtained results and conclude the paper.

\vspace{-0.7cm}

\section{\label{sec:Hamiltonian}Model Hamiltonian and initial conditions}

\begin{figure}[H]
\centering{}\includegraphics[scale=0.4]{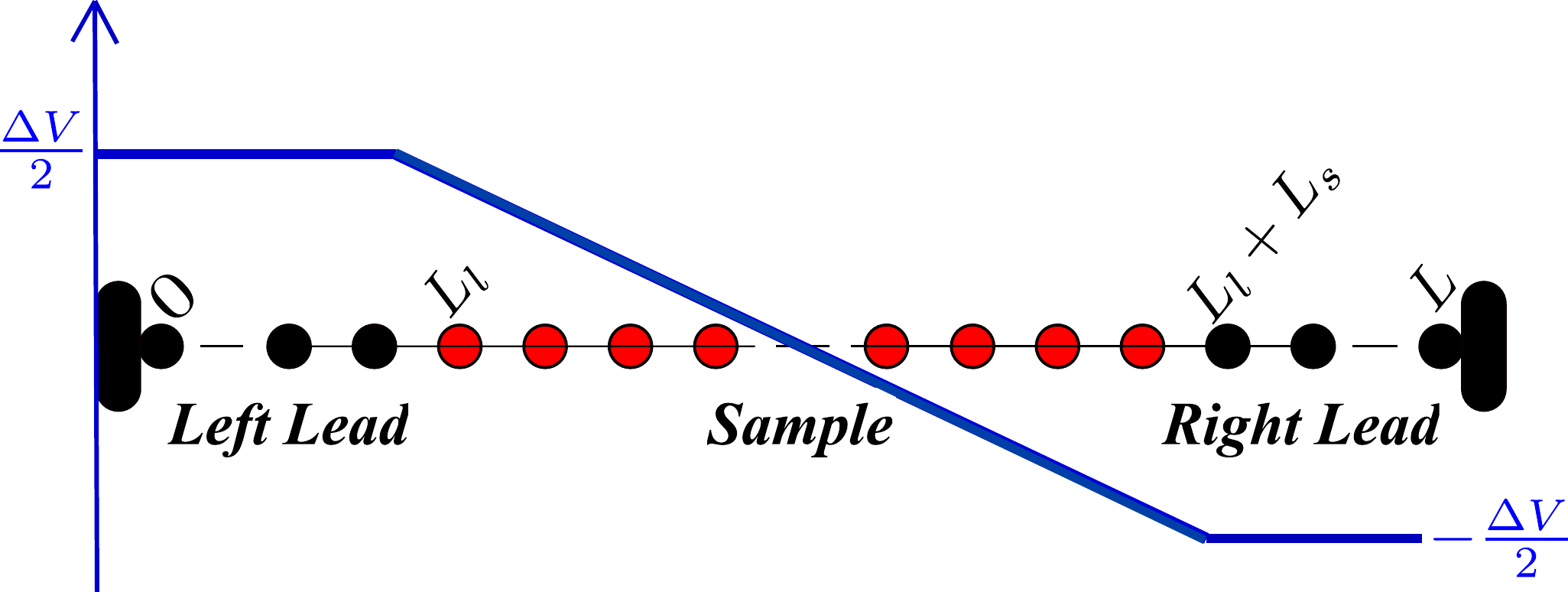}\caption{\label{Scheme_of_the_system}Scheme of the setup used to simulate
the time-dependent LB transport using a one-dimensional sample coupled
to finite leads. The red dots stand for the places where there is
a disordered potential and the blue curve represents the profile of
the externally applied potential. The chain has open boundary conditions.}
\end{figure}

We will study the current dynamics of non-interacting electrons in
a finite one-dimensional tight-binding model, with nearest-neighbor
hoppings. The tight-binding chain is composed by a total of $L$ sites,
with the central $L_{s}$ sites, the sample, having an on-site Anderson
disorder and being subject to a constant electric field. The sites
outside the sample region form the left and right leads {[}each with
$L_{l}=\left(L-L_{s}\right)/2$ sites{]}, are not disordered and hold
a constant electrostatic potential. They will refer to the different
regions in the chain as left lead (LL), sample (S), and right lead
(RL). An illustrative scheme of this setup is shown in Fig.~\ref{Scheme_of_the_system}.
For times $t>0$, the dynamics of the system is governed by the time-independent
Hamiltonian

\vspace{-0.7cm}

{\small{}
\begin{multline}
\mathcal{H}\left(t>0\right)=\sum_{n=0}^{L-1}\left(\epsilon_{n}^{\text{d}}-ev_{n}^{\text{e}}\right)c_{n}^{\dagger}c_{n}-w\sum_{n=0}^{L-2}\left(c_{n+1}^{\dagger}c_{n}+c_{n}^{\dagger}c_{n+1}\right),\label{eq:time-evolution_Hamiltonian}
\end{multline}
}where $c_{n}^{\dagger}$$\left(c_{n}\right)$ are creation (annihilation)
operators for an electron at the chain site $n$, $w$ is the nearest-neighbor
hopping amplitude, $e>0$ is the fundamental charge and $v_{n}^{\text{e}}$
is the electrostatic potential. According to the previous dis-

\vspace{-0.3cm}

\vspace{-0.4cm}

\noindent -cussion $v_{n}^{\text{e}}$ has the form
\begin{equation}
v_{n}^{\text{e}}=\begin{cases}
\frac{\Delta V}{2} & ,\,n\in0,...,L_{l}-1\\
\left(\frac{1}{2}-\frac{n-L_{l}+1}{L_{s}+1}\right)\Delta V & ,\,L_{l}\leq n<L_{l}+L_{s}\\
-\frac{\Delta V}{2} & ,\,n\in L_{l}+L_{s},...,L
\end{cases},\label{PotentialProfile}
\end{equation}
where $\Delta V$ is the applied potential bias, and $\epsilon_{n}^{\text{d}}$
is the Anderson on-site potential disorder, which is only present
in the sample sites, and are taken as random numbers uniformly distributed
inside $\left[-\frac{W}{2},\frac{W}{2}\right]$.

We will study the current dynamics in this system both in the \textit{partitioned}
and \textit{partition-free} approaches. In both cases, the dynamics
for $t>0$ are governed by the Hamiltonian of Eq.~(\ref{eq:time-evolution_Hamiltonian}),
with only the initial state being different.

In the \textit{partitioned approach,} the initial state is formed
by occupied states for the partitioned system, with the bias already
applied. The \textit{partitioned system }is described by the Hamiltonian,

\vspace{-0.3cm}

\begin{equation}
\mathcal{H}^{\text{P}}(t=0)=\mathcal{H}_{\text{LL}}^{\text{P}}+\mathcal{H}_{\text{S}}^{\text{P}}+\mathcal{H}_{\text{RL}}^{\text{P}},\label{eq:partitioned_Hamiltonian}
\end{equation}
with $\mathcal{H}_{\text{LL}}$, $\mathcal{H}_{\text{S}}$, and $\mathcal{H}_{\text{RL}}$
the Hamiltonians, respectively, for the decoupled left lead, sample,
and right lead, to wit
\begin{align}
\mathcal{H}_{\text{LL}}^{\text{P}} & =\sum_{n=0}^{L_{l}-1}\left(-ev_{n}^{\text{e}}\right)c_{n}^{\dagger}c_{n}-w\sum_{n=0}^{L_{l}-2}\left(c_{n+1}^{\dagger}c_{n}+\text{H.c.}\right),\label{eq:partitioned_Hamiltonian_LL}\\
\mathcal{H}_{\text{S}}^{\text{P}} & =\;\qquad\mathllap{\sum_{n=L_{l}}^{L_{l}+L_{s}-1}}\qquad\quad\;\;\mathllap{\left(\epsilon_{n}^{\text{d}}-ev_{n}^{\text{e}}\right)}c_{n}^{\dagger}c_{n}-w\qquad\;\;\mathllap{\sum_{n=L_{l}}^{L_{l}+L_{s}-2}}\qquad\qquad\;\qquad\mathllap{\left(c_{n+1}^{\dagger}c_{n}+\text{H.c.}\right)},\label{eq:partitioned_Hamiltonian_S}\\
\mathcal{H}_{\text{RL}}^{\text{P}} & =\;\;\;\qquad\mathllap{\sum_{n=L_{l}+L_{s}}^{L-1}}\qquad\mathllap{\left(-ev_{n}^{\text{e}}\right)}c_{n}^{\dagger}c_{n}-w\qquad\;\;\mathllap{\sum_{n=L_{l}+L_{s}}^{L-2}}\qquad\qquad\;\qquad\mathllap{\left(c_{n+1}^{\dagger}c_{n}+\text{H.c.}\right)}.\label{eq:partitioned_Hamiltonian_RL}
\end{align}
The initial occupation of the single-electron states is determined
by the independent Fermi levels for each region. Hence, we write $\varepsilon_{\text{F},\text{LL}}=\varepsilon_{\text{F}}+\nicefrac{\Delta V}{2}$,
$\varepsilon_{\text{F},\text{S}}=\varepsilon_{\text{F}}$ and $\varepsilon_{\text{F},\text{RL}}=\varepsilon_{\text{F}}-\nicefrac{\Delta V}{2}$,
as the chemical potential for the left lead, central sample, and right
lead, respectively. $\varepsilon_{\text{F}}$ is a reference chemical
potential. The initial state is thus described by the reduced density
matrix

\vspace{-0.3cm}
\begin{equation}
\rho^{\text{P}}(t=0)=\sum_{r=\text{LL},\text{S},\text{RL}}\sum_{\alpha_{r}}f_{r,\alpha_{r}}^{\text{P}}\left|\Psi_{r,\alpha_{r}}^{\text{P}}\right\rangle \left\langle \Psi_{r,\alpha_{r}}^{\text{P}}\right|,\label{eq:initial_condition_partitioned}
\end{equation}
where $\left|\Psi_{r,\alpha_{r}}^{\text{P}}\right\rangle $ are the
independent single-electron eigenstates of the initial partitioned
Hamiltonian belonging to region $r$, $\mathcal{H}_{r}^{\text{P}}$,
with an energy $\varepsilon_{r,\alpha_{r}}^{\text{P}}$. At any temperature,
the initial occupation of the states is given by the factor $f_{r,\alpha_{r}}^{\text{P}}=f\left(\varepsilon_{r,\alpha_{r}}^{\text{P}}-\varepsilon_{\text{F},r}\right)$,
with $r=\text{LL},\text{S},\text{RL}$ and $f(\varepsilon)=\left(e^{\beta\varepsilon}+1\right)^{-1}$
being the Fermi-Dirac distribution function. Throughout this work,
we will restrict ourselves to the $T=0$ case, where $f(\varepsilon)=\Theta\left(-\varepsilon\right)$
and $\Theta\left(x\right)$ being the usual Heaviside step function.
The hoppings between the leads and the sample are then suddenly switched
on and the time evolution of these states is generated by the Hamiltonian
in Eq. (\ref{eq:time-evolution_Hamiltonian}).

In the \textit{partition-free approach}, the contact between the sample
and the leads is already established in the initial state, but the
bias is not yet applied. Therefore, the initial condition is determined
by populating the eigenstates of the \textit{partition-free} Hamiltonian{\small{}
\begin{equation}
\mathcal{H}^{\text{PF}}(t=0)=\sum_{n=0}^{L-1}\epsilon_{n}^{\text{d}}c_{n}^{\dagger}c_{n}-w\sum_{n=0}^{L-2}\left(c_{n+1}^{\dagger}c_{n}+\text{H.c.}\right),\label{eq:partition-free_Hamiltonian}
\end{equation}
}up to a commonly defined Fermi energy, $\varepsilon_{\text{F}}$.
This initial state is thus described by the reduced density matrix{\small{}
\begin{equation}
\rho^{\text{PF}}(t=0)=\sum_{\alpha}f_{\alpha}^{\text{PF}}\left|\Psi_{\alpha}^{\text{PF}}\right\rangle \left\langle \Psi_{\alpha}^{\text{PF}}\right|,\label{eq:initial_condition_partition-free}
\end{equation}
}with $\left|\Psi_{\alpha}^{\text{PF}}\right\rangle $ being eigenstates
of Eq. (\ref{eq:partition-free_Hamiltonian}) with an eigenenergy
$\varepsilon_{\alpha}^{\text{PF}}$. The initial occupation factor
of the states is similarly given by $f_{\alpha}^{\text{PF}}=f\left(\varepsilon_{\alpha}^{\text{PF}}-\varepsilon_{\text{F}}\right)$.
In this case, the sudden perturbation driving the current is the connection
of the bias potential, $v_{n}^{\text{e}}$, at $t=0$, after which
the time evolution is governed by the Hamiltonian of Eq. (\ref{eq:time-evolution_Hamiltonian}).

We end this section, by noting that the charge current flowing from
site $n$ to site $n+1$, for the Hamiltonian of Eq. (\ref{eq:time-evolution_Hamiltonian}),
is given by

\vspace{-0.2cm}{\small{}
\begin{equation}
\mathcal{I}^{n}=\frac{ew}{i\hbar}\left(c_{n+1}^{\dagger}c_{n}-c_{n}^{\dagger}c_{n+1}\right).\label{eq:local_current}
\end{equation}
}{\small\par}

\vspace{-0.3cm}

\section{\label{sec:Numerical-methods}Numerical methods for current evaluation}

\subsection{\label{subsec:Method_TimeEvolution}Time-resolved current from quantum
evolution of eigenstates}

The dynamics of the system, in the partitioned approach after suddenly
switching on the lead-sample hoppings, or in the partition-free approach
after suddenly switching on the external bias, is governed by the
Hamiltonian (\ref{eq:time-evolution_Hamiltonian}). Therefore, in
both approaches and for $t>0$, the reduced density matrix of the
system evolves according to

\vspace{-0.7cm}

\begin{equation}
i\hbar\frac{d\rho(t)}{dt}=\left[\mathcal{H}(t>0),\rho(t)\right].\label{eq:RDM_eom}
\end{equation}
The solution for this equation, with initial condition given by either
Eq.~(\ref{eq:initial_condition_partitioned}) or (\ref{eq:initial_condition_partition-free}),
is given by

\vspace{-0.5cm}

\begin{equation}
\rho(t)=\sum_{\alpha}f_{\alpha}\left|\Psi_{\alpha}(t)\right\rangle \left\langle \Psi_{\alpha}(t)\right|,
\end{equation}
with the single-electron states evolving according to Eq.~(\ref{eq:time-evolution_Hamiltonian}):
$\left|\Psi_{\alpha}(t)\right\rangle =e^{-\frac{i}{\hbar}\mathcal{H}(t>0)t}\left|\Psi_{\alpha}\right\rangle $,
with $\left|\Psi_{\alpha}\right\rangle $ being the single-electron
eigenstates of either $\mathcal{H}^{\text{P}}(t=0)$ or $\mathcal{H}^{\text{PF}}(t=0)$
(with occupation $f_{\alpha}$), for the \textit{partitioned} and
\textit{partition-free} approaches, respectively.

The expected value of the current, as a function of time, is given
by

\vspace{-0.6cm}
\begin{multline}
I^{n}(t)=\frac{ew}{i\hbar}\sum_{\alpha\in\text{occupied}}\left(\left\langle \left.\Psi_{\alpha}(t)\right|n+1\right\rangle \left\langle n\left|\Psi_{\alpha}(t)\right.\right\rangle \right.\\
\left.-\left\langle \left.\Psi_{\alpha}(t)\right|n\right\rangle \left\langle n+1\left|\Psi_{\alpha}(t)\right.\right\rangle \right),\label{eq:LocalCurrent_withSPES}
\end{multline}
where $\left|n\right\rangle $ represents the state localized at site
$n$. We also used the fact that at $t=0$, $f_{\alpha}=1$ for initially
occupied states and $f_{\alpha}=0$ for empty states. The above expression,
allows us to evaluate the current flowing from site $n$ to site $n+1$
provided we know the time evolution of the initial single-electron
states. Although correct, Eq. (\ref{eq:LocalCurrent_withSPES}) is
not very convenient from a numerical point of view, since for each
initially occupied state, we would have to perform one time-evolution.
A more convenient expression is obtained by writing $\left\langle \left.\Psi_{\alpha}(t)\right|n\right\rangle =\left\langle \Psi_{\alpha}\right|e^{\frac{i}{\hbar}\mathcal{H}\left(t>0\right)t}\left|n\right\rangle =\left\langle \left.\Psi_{\alpha}\right|n(-t)\right\rangle $,
such that instead of evolving the initial eigenstate forwards in time,
we evolve the localize states backwards in time.\footnote{Note that, despite being an evolution for negative times, the time-evolution
operator used to do it is the one which includes the external perturbation:
lead-sample hopping in the partitioned approach, and bias in the partition-free
approach.} 

The current can therefore be expressed as

\begin{equation}
I^{n}(t)=\frac{2ew}{\hbar}\text{Im}\sum_{\alpha}\left\langle n(-t)\left|\Psi_{\alpha}\right.\right\rangle \left\langle \left.\Psi_{\alpha}\right|n+1(-t)\right\rangle .
\end{equation}

Despite being equivalent to Eq.~(\ref{eq:LocalCurrent_withSPES}),
this last expression allows for a great gain in computational efficiency,
as the the number of required time evolutions is reduced from $\mathcal{O}\left(L\right)$
to only two, for each single-time calculation of the current between
sites $n$ and $n+1$.

Numerically, the time-evolution of the localized states $\left|n\right\rangle $,
for very large chains, is computed efficiently using a polynomial
Chebyshev expansion~\citep{KPM_Timeevolution_Talezer1984,fehske_numerical_2009}
of the time evolution operator $\mathcal{U}(t)=e^{-\frac{i}{\hbar}\mathcal{H}t}$
(for details, see Appendix \ref{appx:TimeEvolutionMethod}). Finally,
the single-electron eigenstates of $\mathcal{H}\left(t=0\right)$
were calculated ``on the fly'' using a memory-efficient algorithm
developed by Fernando~\citep{fernando_computing_1997}.

\subsection{\label{subsec:Steady-state-Landauer}Landauer-Büttiker formula for
the steady-state current}

In Sec. \ref{sec:NumericalResults}, the time-resolved current across
the sample with finite leads, calculated using the quantum evolution
of the occupied states, will be compared with the steady-state value
for the current as given by the Landauer-Büttiker formula for the
same sample attached to infinite leads. We evaluate this current using
the Caroli-Meir-Wingreen form of the Landauer-Büttiker formula\,\citep{caroli_direct_1971,meir_landauer_1992},
which for a two-terminal device, at zero temperature, reads as 

\vspace{-0.7cm}

\begin{equation}
I_{\text{LB}}=\frac{e}{2\pi\hbar}\int_{\varepsilon_{\text{F}}-e\frac{\Delta V}{2}}^{\varepsilon_{\text{F}}+e\frac{\Delta V}{2}}d\varepsilon T\left(\varepsilon\right),\label{eq:Mier-WingreenFormula-1}
\end{equation}

\noindent where the energy dependent transmission function is expressed
in terms of Green's functions as

\vspace{-0.3cm}

\begin{equation}
T(\varepsilon)=\text{Tr}\left[\bm{G}^{A}(\varepsilon)\cdot\bm{\Gamma}_{\text{RL}}(\varepsilon)\cdot\bm{G}^{R}(\varepsilon)\cdot\bm{\Gamma}_{\text{LL}}(\varepsilon)\right],
\end{equation}
where $\bm{\Gamma}_{\text{LL}/\text{RL}}(\varepsilon)$ are real-space
spectral functions of the unattached leads and $\bm{G}^{R/A}(\varepsilon)$
are the real-space retarded/advanced Green's function of the central
sample, in the presence of the leads. For our particular one-dimensional
model, the leads' spectral functions are matrices with the only non-zero
elements between boundary sites, i.e., $\Gamma_{\text{LL}}^{L_{l}-1,L_{l}-1}(\varepsilon)=\Gamma_{\text{LL}}(\varepsilon)$
and $\Gamma_{\text{RL}}^{L_{l}+L_{s},L_{l}+L_{s}}(\varepsilon)=\Gamma_{\text{LL}}(\varepsilon)$
, where $\Gamma_{\text{LL}/\text{RL}}(\varepsilon)=w^{2}\rho_{\text{LL}/\text{RL}}(\varepsilon)$.
The functions $\rho_{\text{LL}/\text{RL}}(\varepsilon)$ are surface
density of states of the leads which may be computed analytically
yielding:

\vspace{-0.7cm}

\begin{multline}
\rho_{\text{LL}/\text{RL}}(\varepsilon)=\Theta\left(4w^{2}-\left(\varepsilon\mp\frac{e\Delta V}{2}\right)^{2}\right)\\
\times\frac{1}{w^{2}}\sqrt{4w^{2}-\left(\varepsilon\mp\frac{e\Delta V}{2}\right)^{2}}.\label{eq:Gamma-1}
\end{multline}
Therefore, the final form for the transmission function reads

\begin{equation}
T(\varepsilon)=w^{4}\rho_{\text{LL}}(\varepsilon)\rho_{\text{RL}}(\varepsilon)\left|G_{L_{l}-1,L_{l}+L_{s}}^{R}(\varepsilon)\right|^{2}.\label{eq:LandauerFormula1D}
\end{equation}
For each central sample, the retarded Green's function in Eq.~(\ref{eq:LandauerFormula1D})
was calculated by using the well-known \emph{recursive Green's function
method} \citep{MacKinnon1985,wimmer_quantum_2009,lewenkopf_recursive_2013},
using the surface Green function of the semi-infinite one-dimensional
leads as boundary conditions, as detailed in Appendix~\ref{appx:RecursiveTransferMatrixAppendix}.

\section{\label{sec:NumericalResults}Numerical results and comparison with
the Landauer formula}

We evaluated the time-dependent current in finite open chains, using
the method described in Sec.~\ref{subsec:Method_TimeEvolution},
for both clean and disordered samples and considering both the \textit{partitioned}
and \textit{partition-free} initial conditions. This current was then
compared with the Landauer expression for the steady-state current
flowing through the same sample attached to infinite leads, as described
in Sec.~\ref{subsec:Steady-state-Landauer}.

Our results are summarized in Fig.~(\ref{fig:PartitionedVsNonPartitioned}).
For both initial conditions, three transport regimes are clearly distinguished
for large enough leads:\textit{ (i) }initially, we have a transient
regime up to a time $t_{\text{stab}}$, after which \textit{(ii)}
the current tends to an approximately constant quasi-steady-state
value, which last up to \textit{(iii)} a recurrence time, $t_{\text{r}}$,
after which an inversion of the current occurs. These three transport
regimes have been previously reported for non-interacting fermions
in systems with finite leads, in Refs.~\citep{bushong_approach_2005,chien_bosonic_2012},
with an initial state where the leads are connected to the device,
but one of the leads is depleted. These transport regimes have also
been observed for interacting fermions (at a TDDFT-ALDA level) with
finite leads by Bushong \textit{et al}~\citep{bushong_approach_2005}.
These three regimes have also been discussed in Di Ventra \textit{et
al}~\citep{ventra_transport_2004} and Pal \textit{et al} \citep{pal_emergence_2018}.
However, to the best of our knowledge the effect of different initial
conditions, in each of the transport regimes has never been explored.

In the following, we will analyze in detail each of these regimes,
analyzing how the different time-scales depend on the variables of
the problem.
\begin{widetext}
\onecolumngrid

\begin{figure}[H]
\begin{centering}
\includegraphics[scale=0.37]{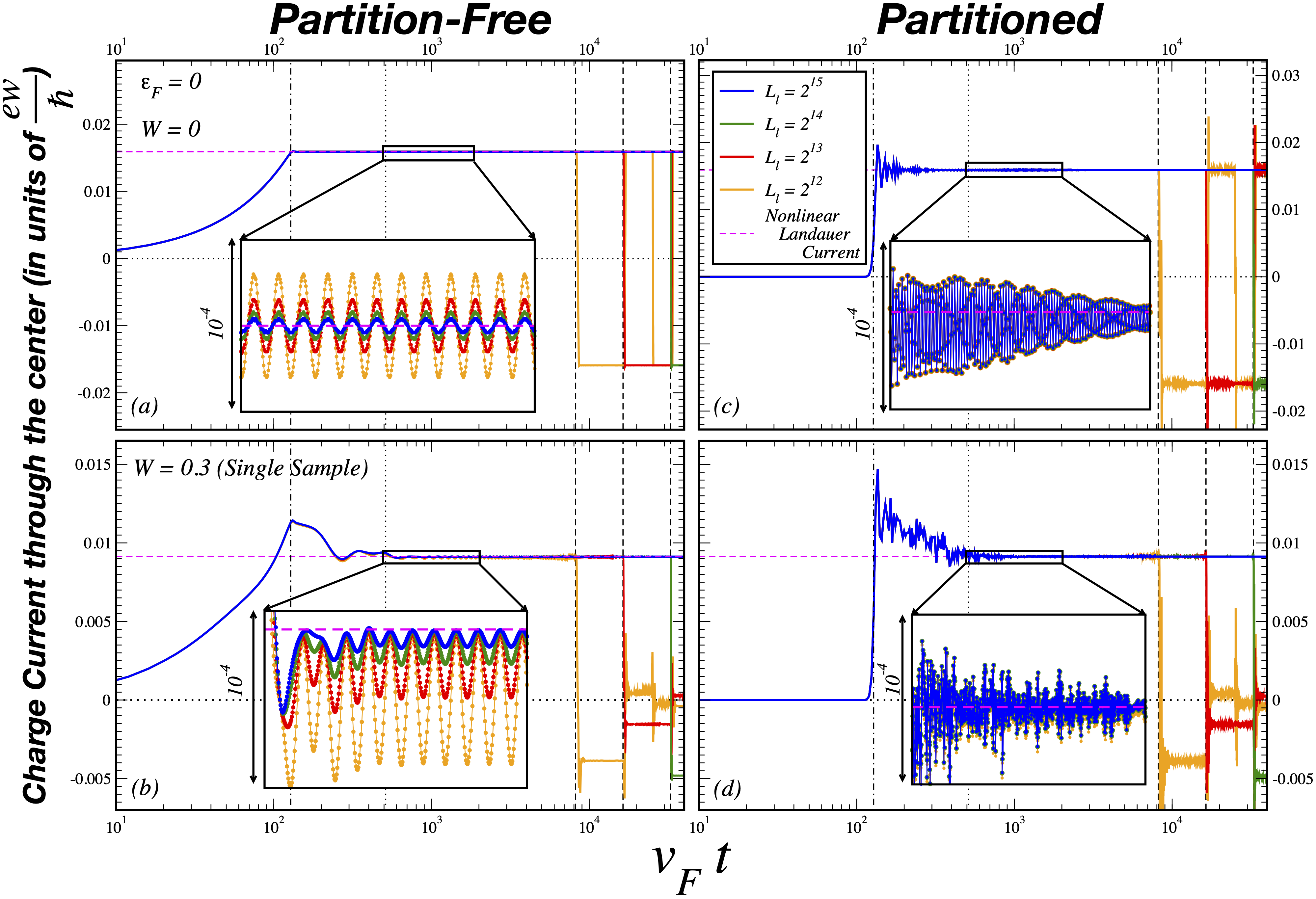}
\par\end{centering}
\caption{\label{fig:PartitionedVsNonPartitioned}Comparison of the time-dependent
current traversing the center of a sample with $L_{s}=256$ sites,
obtained in both the \textit{partition-free }{[}\textit{(a)} and\textit{
(b)}{]} and \textit{partitioned }approaches {[}\textit{(c)} and\textit{
(d)}{]}, with the steady-state current obtained from the Landauer
formula with semi-infinite leads (dashed magenta lines). In both approaches,
the current is shown for different lead sizes, $\varepsilon_{\text{F}}=0$
and a bias of $\Delta V=0.1w$, without {[}\textit{(a)} and \textit{(c)}{]}
and with disorder {[}\textit{(b)} and \textit{(d)}{]}. The insets
represent the zooms of the current in the quasi-steady-state regime,
with a linear scale in the $x$-axis. As can be seen, the superposed
finite-size oscillations have an amplitude which is less than $1\%$
of the Landauer current in both approaches, although their nature
is different. In the \textit{partition-free} approach there is a decrease
of their amplitude with $L_{l}$, while in the \textit{partitioned}
approach, they die-out only as $t\to+\infty$. In all four panels,
the quasi-steady-state regime is limited by the recurrence time $t_{\text{r}}=2L_{l}/v_{\text{F}}$
(vertical dashed lines). The two transient time scales, the build-up
time $t_{\text{b}}=L_{\text{meas}}/v_{F}$ (dotted vertical lines),
with $L_{\text{meas}}=128$, and $t_{\text{stab}}=2L_{s}/v_{\text{F}}$
(dashed-dotted vertical lines) are also represented in the plots.}
\end{figure}

\twocolumngrid
\end{widetext}

\subsection{Transient Behavior and Stabilization Times}

As one could expect, the transient behavior depends on the initial
preparation of the system, being different for the \textit{partitioned}
and \textit{partition-free} approaches, as evident in Fig. \ref{fig:PartitionedVsNonPartitioned}.

\begin{figure}[H]
\begin{centering}
\includegraphics[width=8.5cm]{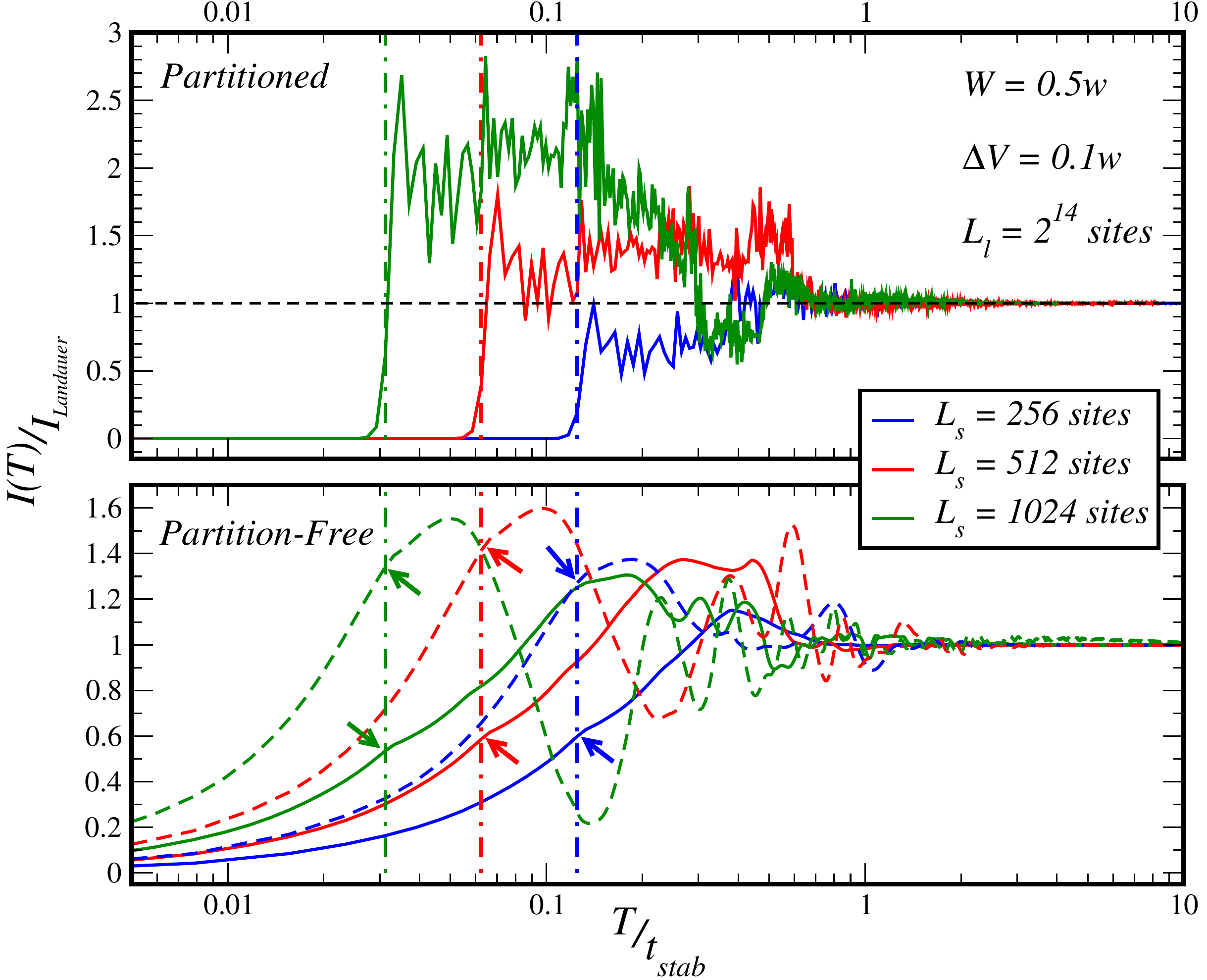}
\par\end{centering}
\caption{\label{fig:Transient}Plots of the normalized time-dependent current
at the $64^{\text{th}}$ hopping, $L_{\text{meas, L}}=L_{\text{meas}}=64$,
of a disordered central sample, calculated using the unitary quantum
dynamics method in the \textit{partitioned }(upper panel) and \textit{partition-free
}approach (lower panel) for a bias of $\Delta V=0.1w$ and different
sample sizes. The time coordinate is rescaled by the stabilization
time-scale, i.e., $t_{\text{stab}}=2L_{\text{s}}/v_{\text{F}}$, which
turns the onset time of the quasi-steady-state roughly independent
of the sample's size in both approaches. The vertical pointed lines
mark the time taken for a Fermi energy state to propagate from the
left lead to the point where the current is being measured, i.e.,
$T=t_{\text{b}}/t_{\text{stab}}$, where the colored arrows highlight
the inflection which occurs at this point for all the curves in \textit{partition-free}
case. Notice that even though $L_{\text{meas}}$, and consequently
$t_{\text{b}}$, is the same for all curves, the vertical lines are
shifted due to the re-scaling by $t_{\text{stab}}$, which depends
on $L_{s}$. In the lower panel, the dashed curves correspond to current
evaluated with a different central disorder configuration.}
\end{figure}

\begin{figure}[H]
\begin{centering}
\includegraphics[width=8.4cm]{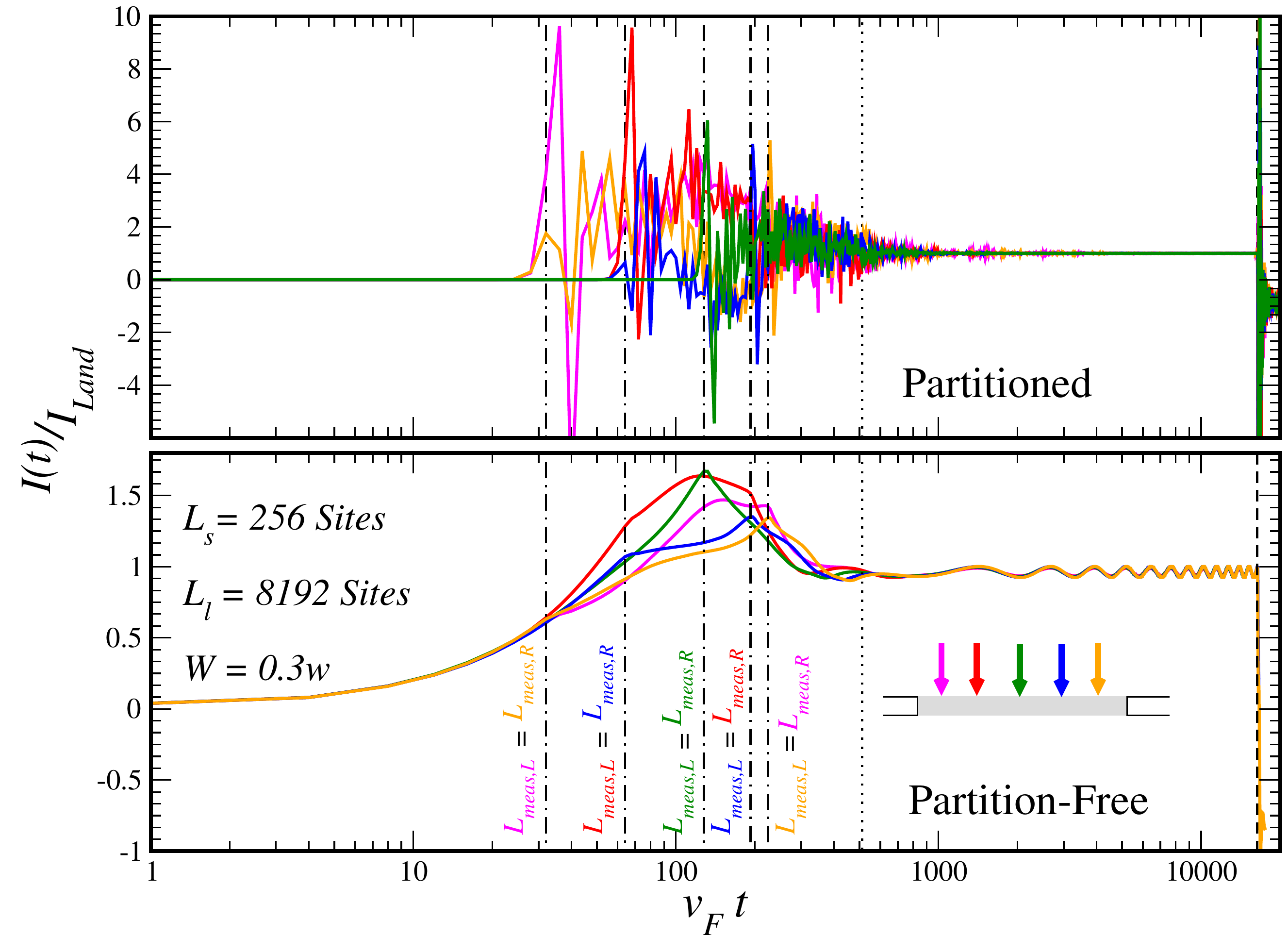}
\par\end{centering}
\caption{\label{fig:Different sites}Current as a function of time measured
at different points of a disorder sample. The current is measured
at the distances $L_{s}/8$, $L_{s}/4$, $L_{s}/2$, $3L_{s}/4$,
and $7L_{s}/8$ from the left lead, and are indicated in the inset
of the bottom panel. The top panel depicts the current for the\textit{
partitioned} setup, while in the bottom panel the \textit{partition-free}
configuration was used. The dashed vertical line represents $t_{\text{r}}$
and the dotted line $t_{\text{stab}}$. The dashed-dotted vertical
lines mark the times $t_{\text{b}}^{\text{L}}$ and $t_{\text{b}}^{\text{R}}$
for the different measuring points. The sample has $L_{s}=256$ sites
and disorder strength of $W=0.3w$. The leads have $L_{l}=2^{13}$
sites and a potential difference of $\Delta V=0.01w$ is applied.}
\end{figure}

In the partitioned case, as can be seen in Figs.~\ref{fig:PartitionedVsNonPartitioned}
(and also in Figs.~ \ref{fig:Transient} and \ref{fig:Different sites}),
the current is initially close to zero up to a build-up time, $t_{\text{b}}$,
after which the current dramatically increases, overshooting the Landauer
current value. We interpret $t_{\text{b}}$ as the time it takes for
a fermion close to the initial reference Fermi level, $\varepsilon_{\text{F}}$,
to travel from the lead-sample boundaries to the hopping where the
current is being probed. We will define $L_{\text{meas, L}}$ and
$L_{\text{meas, R}}$ as the distance from the point where the current
is being measured in the sample to the left and right lead boundaries,
respectively. We will refer to the smallest of these distances as
$L_{\text{meas}}=\min\left(L_{\text{meas, L}},L_{\text{meas, R}}\right)$.
According to this interpretation, the build-up time is given by $t_{\text{b}}=L_{\text{meas}}/v_{\text{F}}$
where $v_{\text{F}}$ is the velocity of a state initially at the
reference Fermi level (the initial Fermi level defined for the sample,
for the partitioned case, and the initial global Fermi level in the
partition-free case). In Fig.\,\ref{fig:Transient}, we show the
current for different samples, measured at $L_{\text{meas, L}}=64$.
The dashed-dotted vertical lines in Figs.~\ref{fig:PartitionedVsNonPartitioned}
and \ref{fig:Transient} indicate the time $t_{\text{b}}=L_{\text{meas}}/v_{\text{F}}$.
The coincidence of these lines with the 

\noindent sharp rise of the current confirms our interpretation. This
is further confirmed in the top panel of Fig.~\ref{fig:Different sites},
where the current measured at different sites is shown as a function
of time, with the dash-dotted vertical lines indicating the times
$t_{\text{b}}^{\text{L/R}}=L_{\text{meas, L/R}}/v_{\text{F}}$. Once
again, we can see that the current in the partitioned case is nearly
zero up to the $t_{\text{b}}=\min\left(t_{\text{b}}^{\text{L}},t_{\text{b}}^{\text{R}}\right)$.

In the \textit{partition-free} setup one observes a gradual increase
of the current from the beginning. As can be seen in the bottom panel
of Fig. \ref{fig:Transient}, the time it takes for a fermion to travel
to the current measuring position, $t_{\text{b}}$, also marks shoulders
in the current (marked by the arrows), a much weaker effect than in
the partitioned case. As a matter of fact, in the partition-free case
weak features (inflection points or peaks) in the current can be observed
at both times $t_{\text{b}}^{\text{L}}$ and $t_{\text{b}}^{\text{R}}$
as is shown in the bottom panel of Fig.~\ref{fig:Different sites}.

After this initial build-up, in both approaches, the current enters
a sample-specific damped oscillatory phase which stabilizes towards
an approximately time-independent value. This stabilization marks
the beginning of the quasi-steady-state regime. As indicated in Fig.~\ref{fig:PartitionedVsNonPartitioned}
and shown by the collapse of the curves in Fig.~\ref{fig:Transient},
for different sample sizes, the quasisteady state exists for times
roughly greater than a stabilization time $t_{\text{stab}}=2L_{s}/v_{\text{F}}$.
Physically, this time can be interpreted as the one needed for a fermion
near the initial reference Fermi level to make a round trip inside
the central sample, thus probing the existing disorder landscape.
The fact that both $t_{\text{b}}$ and $t_{\text{stab}}$ are ballistic
times (i.e., $\propto L_{\text{meas}}$ and $\propto L_{\text{s}}$,
respectively) is consistent with the fact that we are always working
in the ballistic regime of the mesoscopic central sample. Interestingly,
this time-scale is nearly independent of the particular disorder configuration
and the applied bias $\Delta V$.

\vspace{-0.3cm}

\vspace{-0.8cm}

\subsection{Landauer Quasisteady state transport, finite-size effects and recurrence
times}

In both approaches, if the leads are large enough, after the initial
build-up and stabilization of the current, a quasisteady state is
reached for $t>t_{\text{stab}}$, during which the current is approximately
time independent. As the size of the leads increases, the value of
this quasi-steady-state current tends to the sample-specific Landauer
value, independently of the initial preparation of the system (\textit{partitioned}
or \textit{partition-free}). Hence, the present results are numerical
checks to an extension of the memory-loss theorem of Stefanucci \textit{et
al}.~\citep{stefanucci_time-dependent_2004} for the case of finite
leads. The memory-loss theorem \citep{stefanucci_time-dependent_2004}
states that, provided the leads have a continuum spectrum, a steady-state
value of the current is achieved in the $t\rightarrow\infty$ limit,
and that this value is independent of the initial state of the system.

\begin{figure}[H]
\begin{centering}
\includegraphics[width=8.5cm]{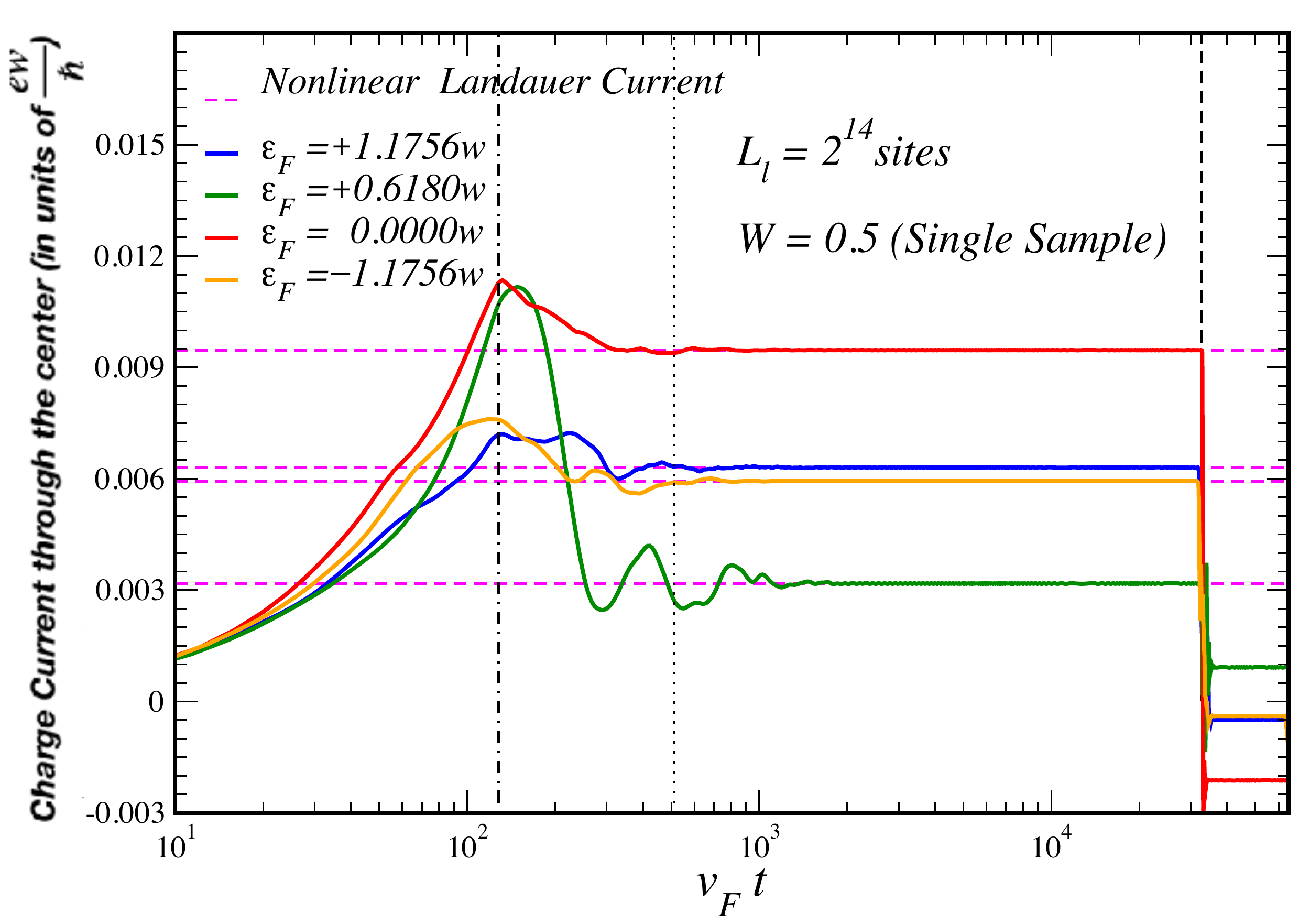}
\par\end{centering}
\caption{\label{fig:DifferentEFs}Plots of the time-dependent current from
the unitary quantum dynamics in the \textit{partition-free} approach
and for a single disordered central sample at different values of
the common reference Fermi energy. The dashed magenta lines correspond,
once again, to the respective steady-state current obtained from the
Landauer formula with semi-infinite leads. The bias used was $\Delta V=0.1w$
and $L_{s}=256$ sites. The black vertical lines have the same meaning
as in Fig. \ref{fig:PartitionedVsNonPartitioned}.}
\end{figure}

As can be seen in Figs.~\ref{fig:PartitionedVsNonPartitioned} and
\ref{fig:DifferentEFs}, in our case, and due to the finite nature
of the leads, which makes their spectrum discrete, a quasi-steady-state
only exists in a finite window of time: $t_{\text{stab}}<t<t_{\text{r}}$.
For $t>t_{\text{r}}$, we observe a drop and inversion of the current.
Similar behavior has also been observed previously \citet{bushong_approach_2005,chien_bosonic_2012}.
Pal\emph{ et al}\,\citep{pal_emergence_2018} pointed out that the
recurrence time, $t_{\text{r}},$ is inversely proportional to the
level spacing of the leads' spectra, which measures how close the
finite leads are to a true continuous spectrum. Our results allow
for an alternative interpretation. As demonstrated in Figs.~\ref{fig:PartitionedVsNonPartitioned}
--- where we show the current for fixed Fermi energy and different
sizes of the leads --- and in Fig.~\ref{fig:DifferentEFs} ---
where we show the current for fixed $L_{l}$ but different Fermi energies
--- the recurrence time is roughly given by $t_{\text{r}}=2L_{l}/v_{\text{F}}$,
where $v_{\text{F}}$ is the Fermi velocity. Notice that $2L_{l}/v_{\text{F}}$
is just the time a fermion close to the Fermi level takes to perform
a round trip inside of a lead, in agreement with what was previously
reported in Ref.~\citep{bushong_approach_2005}. Furthermore, one
also sees that the recursion time is roughly independent of the disorder
on the sample, which is consistent with its previous physical interpretation.

\vspace{-0.3cm}

\vspace{-0.3cm}

One expects that, for sufficiently large leads, the value of the current
in the quasi-steady-state should approach the Landauer value, possibly
with some small oscillations superimposed due to finite size effects
that vanish with increasing lead size. Indeed, this is what we observe
in Fig.~\ref{fig:PartitionedVsNonPartitioned}. However, the way
in which the oscillations vanish depends crucially on the initial
condition of the system. In the \textit{partition-free} approach,
we observe that the current in the quasi-steady-state regime does
not strictly approaches the Landauer value. Instead, there is a small
persistent oscillatory component with a constant amplitude in time,
superposed on its steady-state value, and which persists up to the
recursion time $t_{\textrm{r}}$. Although constant throughout the
quasi-steady-state regime, the amplitude of these oscillations is
seen to decrease as $L_{l}\to\infty$, and the value of the quasi-steady-state
current approaches the Landauer value in a nearly uniform way for
$t_{\text{stab}}<t<t_{\text{r}}$. We also observed, that the period
of these oscillations is roughly inversely proportional to the applied
bias, i.e., $T_{\text{osc}}\propto\Delta V^{-1}$, but does not depend
on either $L_{l}$ or $L_{s}$. Such dependence of the oscillation
period with the applied bias had also been reported by Kurth \textit{et
al}~\citep{Kurth_2005}, although their focus is on systems with
infinite leads. Nevertheless, this behavior gives a hint on the physical
origin of these oscillations. If we consider an occupied eigenstate
$\ket{\Psi(0)}$ of $\mathcal{H}^{\text{PF}}\left(t=0\right)$, having
energy $\varepsilon_{0}$, this can always be written as a linear
combination of eigenstates $\ket{\tilde{\Psi}_{n}}$ of $\mathcal{H}\left(t>0\right)$.
This way, its time-evolved ket, $\ket{\Psi\left(t\right)}$, is simply

\vspace{-0.5cm}

\begin{equation}
\ket{\Psi\left(t\right)}=\sum_{n}e^{-i\hbar^{-1}\varepsilon_{n}t}\braket{\tilde{\Psi}_{n}}{\Psi\left(0\right)}\ket{\tilde{\Psi}_{n}},\label{eq:TimeEvolvedState}
\end{equation}
where $\braket{\tilde{\Psi}_{n}}{\Psi\left(0\right)}$ are the wave-function
overlaps. Since the leads are much larger than the central sample
and the effect of applying a bias is to globally shift the energy
of all Wannier states in the leads to $\pm\Delta V/2$, it is reasonable
to say that the two dominant overlaps in Eq.\,(\ref{eq:TimeEvolvedState})
will be with the states $\ket{\tilde{\Psi}_{\pm}}$ with energies
$\varepsilon_{0}\pm\Delta V/2$. This follows from the fact that these
two are precisely the ones whose real-space wave-functions in either
the right or left lead are stationary waves with the same wavelength
as the original one. Assuming this argument to be true, we may neglect
all the other overlaps in Eq.\,(\ref{eq:TimeEvolvedState}) and write

\vspace{-0.5cm}

\begin{align}
\ket{\Psi\left(t\right)} & \simeq e^{-i\hbar^{-1}\left(\varepsilon_{0}+\frac{\Delta V}{2}\right)t}\braket{\tilde{\Psi}_{+}}{\Psi\left(0\right)}\ket{\tilde{\Psi}_{+}}\nonumber \\
 & +e^{-i\hbar^{-1}\left(\varepsilon_{0}-\frac{\Delta V}{2}\right)t}\braket{\tilde{\Psi}_{-}}{\Psi\left(0\right)}\ket{\tilde{\Psi}_{-}},\label{eq:TimeEvolvedState-1}
\end{align}
which behaves as a two level system, with a Bohr frequency $\omega=\Delta V/\hbar$.
This argument justifies the presence of the time scale associated
to the period of the observed finite-size oscillations. Furthermore,
as $\Delta V$ is reduced, we expect that the overlaps $\braket{\tilde{\Psi}_{\pm}}{\Psi\left(0\right)}$
will increase, thus, we also expect that the amplitude of these oscillations
will increase when $\Delta V$ decreases, a point to which we will
return.

For the \emph{partitioned} approach, a rather different behavior is
observed. In this setup, as can be seen in the insets of Figs.\,\ref{fig:PartitionedVsNonPartitioned}\,(c)
and (d), the amplitude of the oscillations in the quasi-steady-state
decays as time increases, provided $t_{\text{stab}}<t<t_{\text{r}}$.
Furthermore, the amplitude of the oscillations is nearly independent
of the leads' size at any fixed observation time (provided $t<t_{\text{r}}$
for the compared lead sizes). As $L_{l}$ increases, $t_{\text{r}}=2L_{l}/v_{\text{F}}$
also increases and therefore, the oscillations in the quasi-steady-state
will decay for a longer time, thus tending towards the Landauer value
as time tends to $t_{\text{r}},$ $t\rightarrow t_{\text{r}}\rightarrow\infty$.
Note that the physical argument given above in Eqs. (\ref{eq:TimeEvolvedState})
and (\ref{eq:TimeEvolvedState-1}), justifies why one does not observe
persistent oscillations when the system begins in a partitioned setup.

In a true steady state, the value of the current is not only time
independent but must also be position independent, as no charge accumulation
can occur. Hence, we also investigated whether or not this emergent
quasisteady state current in finite chains is homogeneous over the
sample. Indeed, we found out that in the quasisteady state the current
is approximately homogeneous in space, independently of the initial
preparation of the system, for large enough leads and provided we
are far away from the chain's open extremities.\textcolor{blue}{{} }This
observation is exemplified in Fig. \ref{fig:Different sites}, where
we show the time-dependent current for a disordered central sample,
measured at three different bonds: center, left, and right boundaries
of a randomly picked disordered sample. As can be seen, after the
disappearance of the initial transients, the same quasi-steady-state
current is reached at the three positions, apart from the finite-size
oscillations which are out of phase.

\vspace{-0.3cm}

\begin{figure}[H]
\includegraphics[scale=0.34]{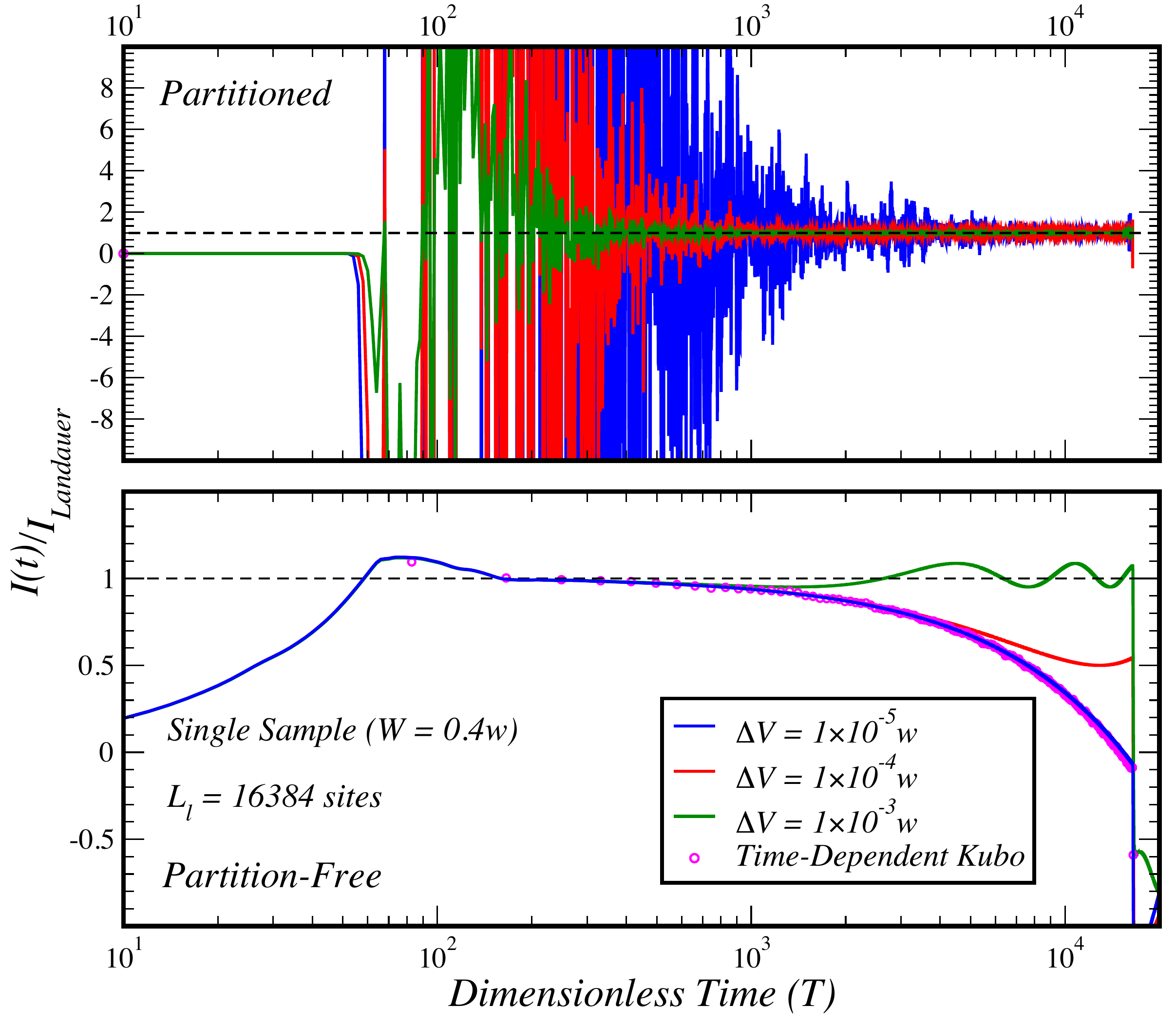}

\caption{\label{fig:LinearResponseExample}Plots of the time-dependent current
across a disordered sample coupled to finite leads with $L_{l}=16\,384$
sites and for different values of $\Delta V\ll w$. The full lines
stand for the results of a fully non-linear calculation using the
quantum dynamics method of last section in the \textit{partitioned}
(upper panel) and \textit{partition-free }approach (lower panel),
while the points stand for the raw evaluation of the linear response
Eq.\,(\ref{eq:LRTCurrent-1}). The last are only present in the \textit{partition-free}
case, where the time-dependent Kubo formula is valid. The value of
the current is normalized to the corresponding Landauer steady-state
value.}
\end{figure}

\vspace{-0.3cm}

Interestingly, the establishment of a well-defined quasisteady state,
for a large but fixed leads size, might not occur for very small biases,
where we would expect linear response theory to hold, depending on
the initial conditions. This is illustrated in Fig. \ref{fig:LinearResponseExample}.
There, we can see that for the \textit{partition-free} setup, no clear
quasi-steady-state is observed for very small biases. This occurs
because, for a fixed lead size and as previously explained, the period
and amplitude (relative to the infinite leads' Landauer value) of
the finite-size oscillations increases with the reduction of the applied
bias. Therefore, for small enough bias, the period of the oscillations
might become larger than the recursion time, and no quasi-steady-state
is observed. A well developed quasi-steady current only emerges provided
$T_{\text{osc}}\ll t_{\text{r}}$. In the \textit{partitioned} setup,
the situation is a bit different and for large times: the current
always tends to the Landauer value with the amplitude of the finite-size
oscillations decreasing over time. These observations seem to be in
agreement with the interpretation of Bushong \textit{et al.} \citep{bushong_approach_2005},
where it is put forward that the observation of a quasi-steady-state
requires the change in the initial spread of the electrons' momenta.
In this reference, this occurs either due to a geometrical constriction
at the lead-sample contact, or due to an initial applied energy barrier.
In our case, it seems that the applied bias is the mechanism by which
electrons change their initial momenta. As a general ``rule-of-thumb'',
we can tell that, in order to observe a quasi-steady-state current
regime with minor finite-size effects, one must always consider biases
that are much larger than the level spacing of the whole system's
spectrum.

\vspace{-0.5cm}

\subsection{Sample-Specific $I-V$ Curves at Large Biases}

We finally point out, that the coincidence between the Landauer value
for the current and the value of the current in the quasisteady state
occurs for any value of the bias potential, as long as it is smaller
than the bandwidth and provided a quasisteady state is established.

\begin{figure}[H]
\begin{centering}
\includegraphics[width=8.3cm]{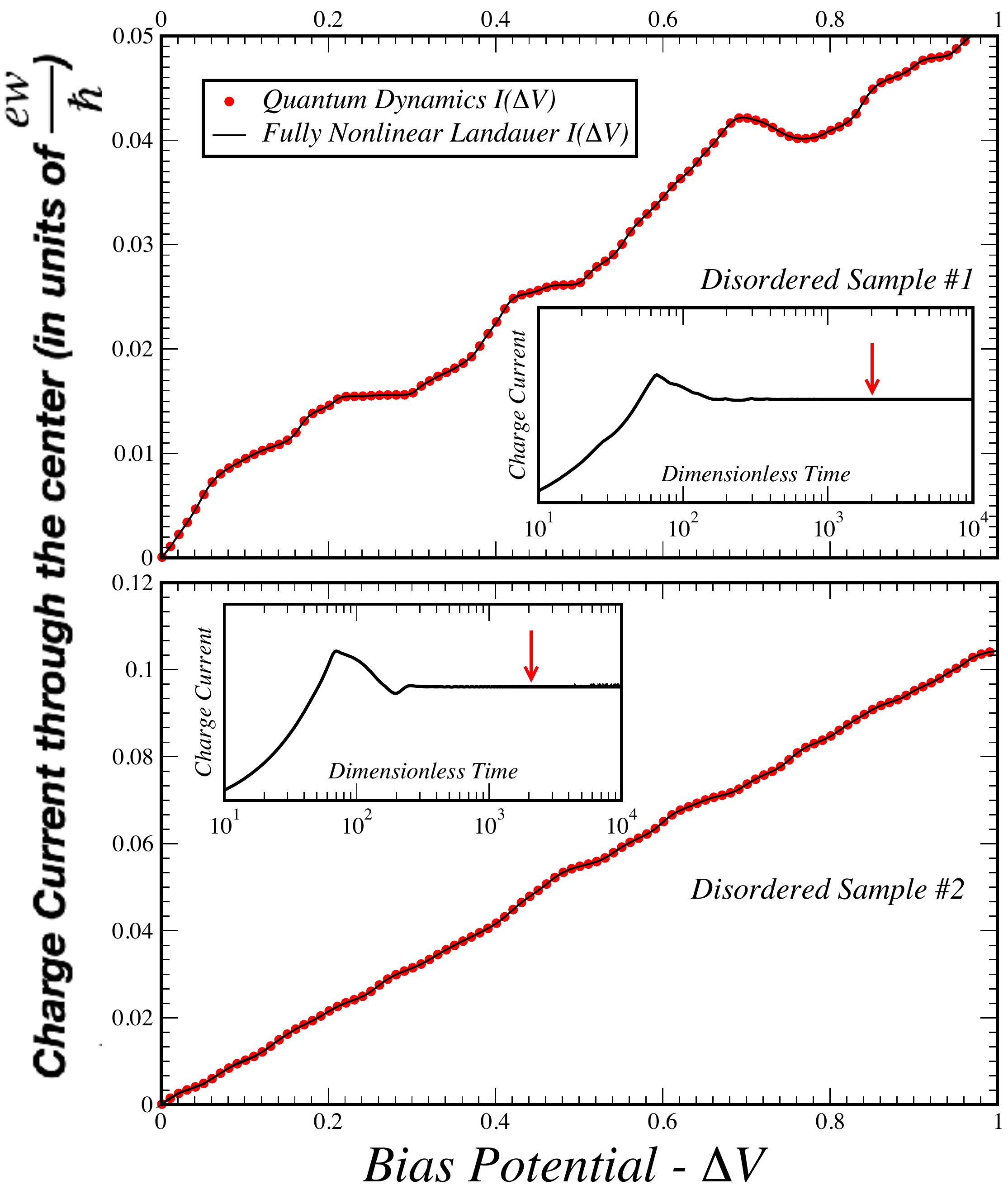}
\par\end{centering}
\caption{\label{fig:IVCurves}Plots of the $I\left(\Delta V\right)$ curves
of two independent disordered samples. The black curves in the main
plots were obtained using the Landauer formula of Eq. (\ref{eq:LandauerFormula1D}).
The red dots were obtained from the quasi-steady-state current of
a quantum dynamics calculation, using the \textit{partition-free}
approach with $L_{l}=2^{14}$ sites. The use of a partitioned approach
could also be done, but would be redundant given that we proved the
numerical equivalence of both approaches in the previous discussion.
In the insets, we highlight with a red arrow the time of measurement
in a plot of $I(t)$. }
\end{figure}

This is illustrated in Fig.~\ref{fig:IVCurves}, where we show values
for the time-dependent current in the quasisteady state regime as
a function of the applied bias, for two random disordered samples,
and compare the results with the value of the Landauer current. The
results clearly confirm that the quasisteady state current seen in
the quantum dynamics calculations with finite leads indeed corresponds
to the Landauer transport predicted for samples coupled to semi-infinite
leads. The agreement between the two approaches was seen to be perfect
for all the range of bias tested and well beyond linear response.

\vspace{-0.3cm}

\section{\label{sec:Emergence-of-Landauer}Emergence of Landauer transport
within linear response in the partition-free approach}

The numerical studies of the previous section show that a quasi-steady-state
transport regime, with an approximately uniform and time-independent
current, emerges across finite systems subjected to a potential bias
and coupled to finite but large leads. Moreover, the results also
show that for large enough leads, the value of this quasi-steady-state
current coincides with the Landauer result for the transport's steady
state with semi-infinite leads. In this section, we will try to shed
further light on these numerical results using a semi-analytical procedure.
In order to make as much analytical progress as possible, we shall
restrict ourselves to the \textit{partition-free} case and small biases,
such that we can study the current using Kubo linear response theory
in the applied bias, $\Delta V$.

\subsection{Time-dependent Kubo formula for a sudden connection}

We will always consider the \textit{partition-free} Hamiltonian at
$t=0$ as the unperturbed Hamiltonian for this case, i.e.,

\vspace{-0.7cm}
\begin{align}
\mathcal{H}_{0} & =\mathcal{H}^{\text{PF}}(t=0)\nonumber \\
 & =\sum_{n=0}^{L-1}\epsilon_{n}^{\text{d}}c_{n}^{\dagger}c_{n}-w\sum_{n=0}^{L-2}\left(c_{n+1}^{\dagger}c_{n}+c_{n}^{\dagger}c_{n+1}\right)\label{eq:unperturbed_Hamiltonian}
\end{align}
and treat the applied potential bias as the current-driving perturbation,

\vspace{-0.8cm}

\begin{equation}
\mathcal{V}(t)=-e\Theta(t)\sum_{n=0}^{L-1}v_{n}^{\text{e}}c_{n}^{\dagger}c_{n}.\label{eq:PerturbationHamiltonian}
\end{equation}
with the electrostatic potential profile, $v_{n}^{\text{e}}$, given
by Eq.~(\ref{PotentialProfile}).

In order to derive a time-dependent Kubo formula for the current,
we will start by writing the equation of motion for the reduced density
matrix, Eq.~(\ref{eq:RDM_eom}), in the eigenbasis of the unperturbed
Hamiltonian. Thus, we obtain
\begin{equation}
\frac{d}{dt}\rho_{\alpha\beta}(t)=-\frac{i}{\hbar}\left(\varepsilon_{\alpha}-\varepsilon_{\beta}\right)\rho_{\alpha\beta}(t)-\frac{i}{\hbar}\left[\mathcal{V}(t),\rho(t)\right]_{\alpha\beta},\label{eq:EOM_RDM_eigenbasis}
\end{equation}
where $O_{\alpha\beta}(t)=\left\langle \Psi_{\alpha}\left|O\right|\Psi_{\beta}\right\rangle $
and $\left|\Psi_{\alpha}\right\rangle $ is an eigenstate of $\mathcal{H}_{0}$
with energy $\varepsilon_{\alpha}$. Within linear response theory,
we write the reduced density matrix as
\begin{equation}
\rho_{\alpha\beta}\left(t\right)=\delta_{\alpha\beta}f\left(\varepsilon_{\alpha}\right)+\delta\rho_{\alpha\beta}\left(t\right),
\end{equation}
where $\rho_{\alpha\beta}\left(0\right)=\delta_{\alpha\beta}f\left(\epsilon_{\alpha}\right)$
is the initial equilibrium reduced density matrix and $\delta\rho_{\alpha\beta}\left(t\right)$
is a small correction, which in linear response is assumed to be $\propto\mathcal{V}(t)$.
Disregarding any contributions of $\mathcal{O}\left(\mathcal{V}^{2}\right)$
in the equation of motion, we obtain

\vspace{-0.7cm}

\begin{multline}
\frac{d}{dt}\delta\rho_{\alpha\beta}(t)=-\frac{i}{\hbar}\left(\varepsilon_{\alpha}-\varepsilon_{\beta}\right)\delta\rho_{\alpha\beta}(t)\\
-\frac{ie}{\hbar}\Theta(t)\Gamma_{\alpha\beta}\left(f(\varepsilon_{\alpha})-f(\varepsilon_{\beta})\right).\label{eq:EMRDM-1-2}
\end{multline}
where $\Gamma_{\alpha\beta}$ are the matrix elements of the applied
potential bias,

\vspace{-0.7cm}
\begin{equation}
\Gamma_{\alpha\beta}=\sum_{n}\psi_{\alpha}^{*}(n)\psi_{\beta}(n)v_{n}^{\text{e}},
\end{equation}
and $\psi_{\alpha}(n)$ is the amplitude of the eigenstate $\ket{\Psi_{\alpha}}$
on site $n$, i.e. $\psi_{\alpha}(n)=\left\langle n\left|\Psi_{\alpha}\right.\right\rangle $.
Now, using the fact that $\delta\rho_{\alpha\beta}(t<0)=0$, it is
possible to integrate Eq.~(\ref{eq:EMRDM-1-2}), obtaining
\begin{equation}
\delta\rho_{\alpha\beta}(t)=-e\Gamma_{\alpha\beta}\frac{\Delta f_{\alpha\beta}}{\Delta\varepsilon_{\alpha\beta}}\left(1-e^{-\frac{i}{\hbar}\Delta\varepsilon_{\alpha\beta}t}\right),\label{eq:EMRDM-1-2-1-1-1-1}
\end{equation}
where $\Delta f_{\alpha\beta}=f(\varepsilon_{\alpha})-f(\varepsilon_{\beta})$
and $\Delta\varepsilon_{\alpha\beta}=\varepsilon_{\alpha}-\varepsilon_{\beta}$.
The expected value of the current that flows from site $n$ to $n+1$,
is thus given by
\begin{equation}
I^{n}\left(t\right)=\frac{ie^{2}w}{\hbar}\sum_{\alpha,\beta}\Pi_{\alpha\beta}^{n}\Gamma_{\beta\alpha}\frac{\Delta f_{\alpha\beta}}{\Delta\varepsilon_{\alpha\beta}}\left(1-e^{-\frac{i}{\hbar}\Delta\varepsilon_{\alpha\beta}t}\right),\label{eq:LRTCurrent}
\end{equation}
where we introduced 
\begin{equation}
\Pi_{\alpha\beta}^{n}=\psi_{\alpha}^{*}(n+1)\psi_{\beta}(n)-\psi_{\alpha}^{*}(n)\psi_{\beta}(n+1),\label{eq:Gamma_definition}
\end{equation}
which are the matrix elements of the local current operator between
sites $n$ and $n+1$, up to a dimension-full multiplicative factor.

By further noticing that the amplitudes $\psi_{\alpha}(n)$ may be
chosen as all real and $\Pi^{s}$ is an anti-symmetric matrix, one
can rewrite Eq.~(\ref{eq:LRTCurrent}) in the following way:

\vspace{-0.4cm}

\begin{equation}
I^{n}\left(t\right)=\frac{2e^{2}w}{\hbar}\qquad\qquad\quad\qquad\qquad\qquad\qquad\qquad\mathllap{\sum_{\overset{\alpha}{\left(\varepsilon_{\alpha}\leq\varepsilon_{F}\right)}}\sum_{\overset{\beta}{\left(\varepsilon_{\beta}>\varepsilon_{F}\right)}}\Pi_{\alpha\beta}^{n}\Gamma_{\alpha\beta}\frac{\sin\left(\frac{\Delta\varepsilon_{\alpha\beta}t}{\hbar}\right)}{\Delta\varepsilon_{\alpha\beta}}.}\label{eq:LRTCurrent-1}
\end{equation}
which is our final time-dependent Kubo formula for the current.

Obviously, one cannot give a general rule for establishing the validity
regime of Eq.\,(\ref{eq:LRTCurrent-1}), since that will depend crucially
on the properties of the central disordered sample. However, for each
sample, there is always a value of $\Delta V$ sufficiently small,
such that a linear response theory for the current is valid. We depict
such an example in the upper panel of Fig.\,\ref{fig:LinearResponseExample},
where the current traversing the central bond of a disordered sample,
as obtained from Eq.\,(\ref{eq:LRTCurrent-1}), is compared with
the one obtained from the fully nonlinear quantum dynamics of sec.\,\ref{sec:Numerical-methods}
in the \textsl{partition-free} approach. As a further short comment
on the plots of Fig.\,\ref{fig:LinearResponseExample}, it is interesting
to note that, for the parameters used, it seems that no quasi-steady-state
plateau emerges from the quantum dynamics close to the linear response
regime. As referred before, this is simply a consequence of a greater
relevance of the finite-size oscillations which, now, have a period
larger than the recurrence time and a much larger relative amplitude.

\subsection{Representation of the eigenstates in terms of the sample's quantum
reflection/transmission coefficients}

In order to make an effective use of Eq.~(\ref{eq:LRTCurrent-1})
and make analytic progress we must be able to find a semi-analytical
expression for the matrix elements $\Pi_{\alpha\beta}^{n}$ and $\Gamma_{\alpha\beta}$,
which, in principle, requires the knowledge of the eigenfunctions
in the whole chain. These wave functions usually present a very complicated
structure inside the disordered central sample, but for large enough
leads, we actually only need to know their form in the leads. On the
one hand, the $\Pi_{\alpha\beta}^{n}$ matrix elements only require
the knowledge of local amplitudes in the two adjacent sites across
which the current is being measured. Hence, we can simply choose to
measure it outside the sample. On the other hand, we expect the current
to be dominated by states that are not localized in the disordered
sample, but instead are delocalized in the leads. Hence, we only need
to calculate the $\Gamma_{\alpha\beta}$ matrix elements between delocalized
states. For such states, and provided the leads are much larger than
the disordered sample region, we can approximate

\begin{equation}
\Gamma_{\alpha\beta}=\sum_{n}\psi_{\alpha}^{*}(n)\psi_{\beta}(n)v_{n}^{\text{e}}\simeq\qquad\mathllap{\sum_{n\in\text{Leads}}}\psi_{\alpha}^{*}(n)\psi_{\beta}(n)v_{n}^{\text{e}}.\label{eq:large_lead:approx}
\end{equation}
This approximation, allows us to evaluate the current $I^{n}\left(t\right)$
in the leads, without knowing the shape of the eigenwave functions
inside the central sample.

Next, we notice that the form of the scattering eigenstates in the
leads can be expressed in terms of the complex reflection and transmission
coefficients of the central sample. For perfect leads, the wave functions
of the eigenstates will have the form of a coherent superposition
of left and right propagating plane-waves. With a change of notation
from the previous section, we will relabel sites of the left lead
with indices $n=-L_{l},...,-1$ and the ones of the right lead with
$n=1,...,L_{l}$. Using this notation, the form of the eigenstate
wave function $\left|\Psi_{k}\right\rangle $ in the leads have the
form

\vspace{-0.7cm}

\begin{multline}
\psi_{k}(n)=\left\langle n\left|\Psi_{k}\right.\right\rangle \\
=\begin{cases}
\Psi_{+}^{L}e^{ik\left(n-1\right)}+\Psi_{-}^{L}e^{-ik\left(n-1\right)}, & \quad\qquad\qquad\mathllap{-L_{l}\leq n\leq-1}\\
\Psi_{+}^{R}e^{ik\left(n-L_{s}\right)}+\Psi_{-}^{R}e^{-ik\left(n-L_{s}\right)}, & 1\leq n\leq L_{l}
\end{cases},\label{eq:wavefunctions_leads}
\end{multline}
being labeled by a crystal momentum $k$, and with $\Psi_{+/-}^{L(R)}$
being the amplitude of a right/left propagating state in the left
(right) lead. Notice that the time-independent Schrödinger equation
inside the leads, still allows us to relate the crystal momentum $k$
to the energy of the state as $E=-2t\cos(k)$, i.e. the same as for
an infinite periodic chain. As usual in one-dimensional scattering
problems, the amplitudes of propagating states on the left and right
leads can be related by a transfer matrix, $\mathcal{M}(k)$:
\begin{equation}
\left(\begin{array}{c}
\Psi_{+}^{R}\\
\Psi_{-}^{R}
\end{array}\right)=\mathcal{M}\left(k\right)\cdot\left(\begin{array}{c}
\Psi_{+}^{L}\\
\Psi_{-}^{L}
\end{array}\right),\label{eq:S_Matrixef-4-1}
\end{equation}
In the presence of time-reversal symmetry, the transfer matrix has
the general form{\small{}
\begin{equation}
\mathcal{M}\left(k\right)=\left(\begin{array}{cc}
\frac{1}{\abs{t\left(k\right)}}e^{i\phi\left(k\right)} & -\frac{\abs{r\left(k\right)}}{\abs{t\left(k\right)}}e^{-i\theta\left(k\right)+i\phi\left(k\right)}\\
-\frac{\abs{r\left(k\right)}}{\abs{t\left(k\right)}}e^{i\theta\left(k\right)-i\phi\left(k\right)} & \frac{1}{\abs{t\left(k\right)}}e^{-i\phi\left(k\right)}
\end{array}\right),
\end{equation}
}where $\abs{t\left(k\right)}/\abs{r\left(k\right)}$ and $\phi\left(k\right)/\theta\left(k\right)$
are the moduli and phases of the transmission and reflection coefficients,
respectively. Moreover, for any sample one has $\det\mathcal{M}=1$,
which implies the conservation of current, i.e., $\abs t^{2}+\abs r^{2}=1$.
These coefficients are physical characteristics of the central sample
only and, thus, may be rightfully calculated by assuming the leads
as semi-infinite. The determination of the reflection and transmission
coefficients of a specific sample, in general, can only be done numerically,
using the method detailed in Appendix \ref{appx:RecursiveTransferMatrixAppendix}.
The great advantage of this method is that, once this calculation
is done, the wave functions in the leads can be expressed in terms
of only a few parameters. Additionally, to obtain the eigenstates,
we must further impose open boundary conditions at the ends of the
leads, i.e.,

\vspace{-0.5cm}
\begin{equation}
\psi_{k}\left(-L_{l}-1\right)=\psi_{k}\left(L_{l}+1\right)=0.\label{eq:boundary_condition}
\end{equation}

\vspace{-0.4cm}

\begin{figure}[H]
\begin{centering}
\includegraphics[width=7.7cm]{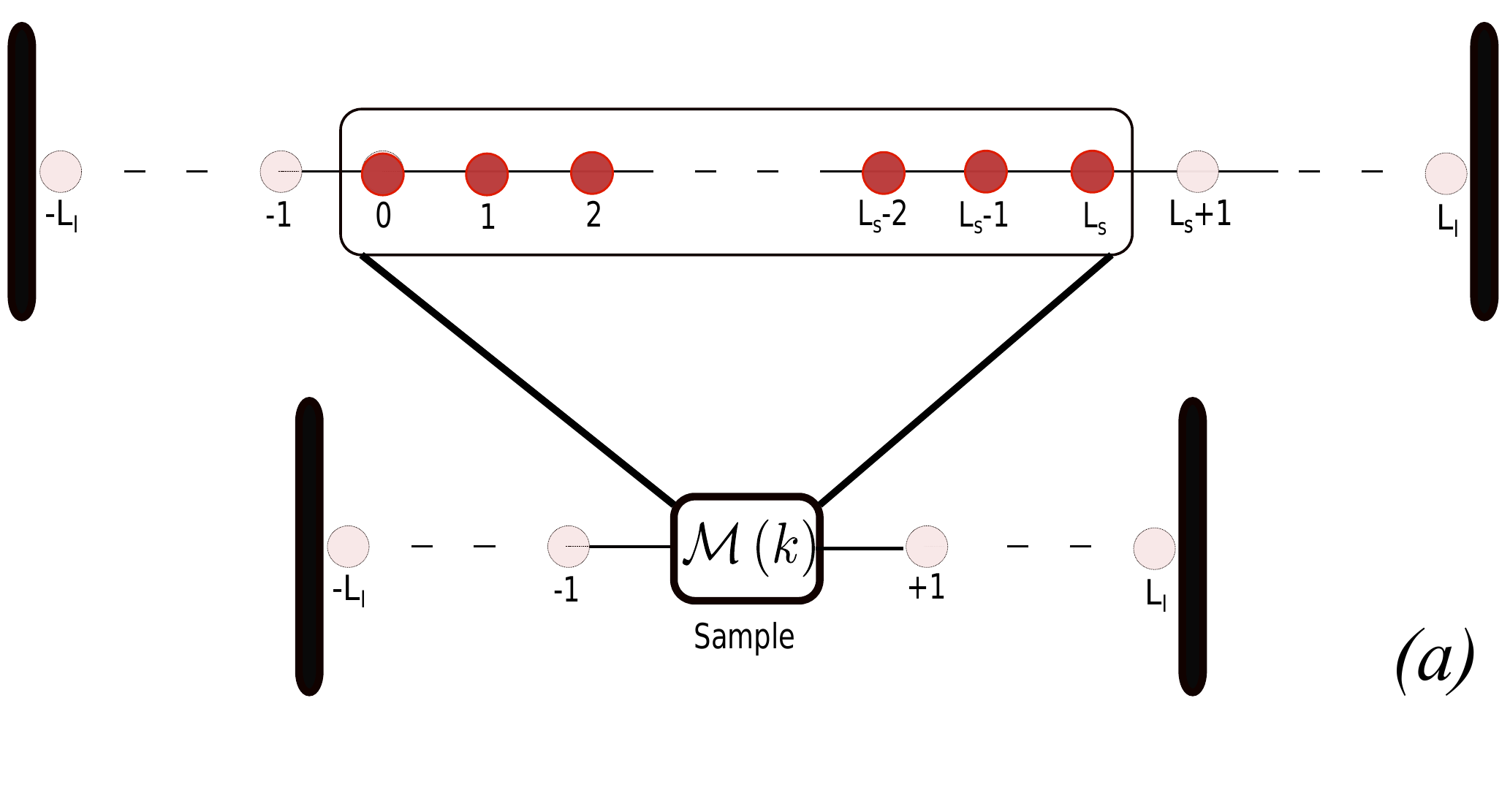}
\par\end{centering}
\begin{centering}
\includegraphics[width=8.2cm]{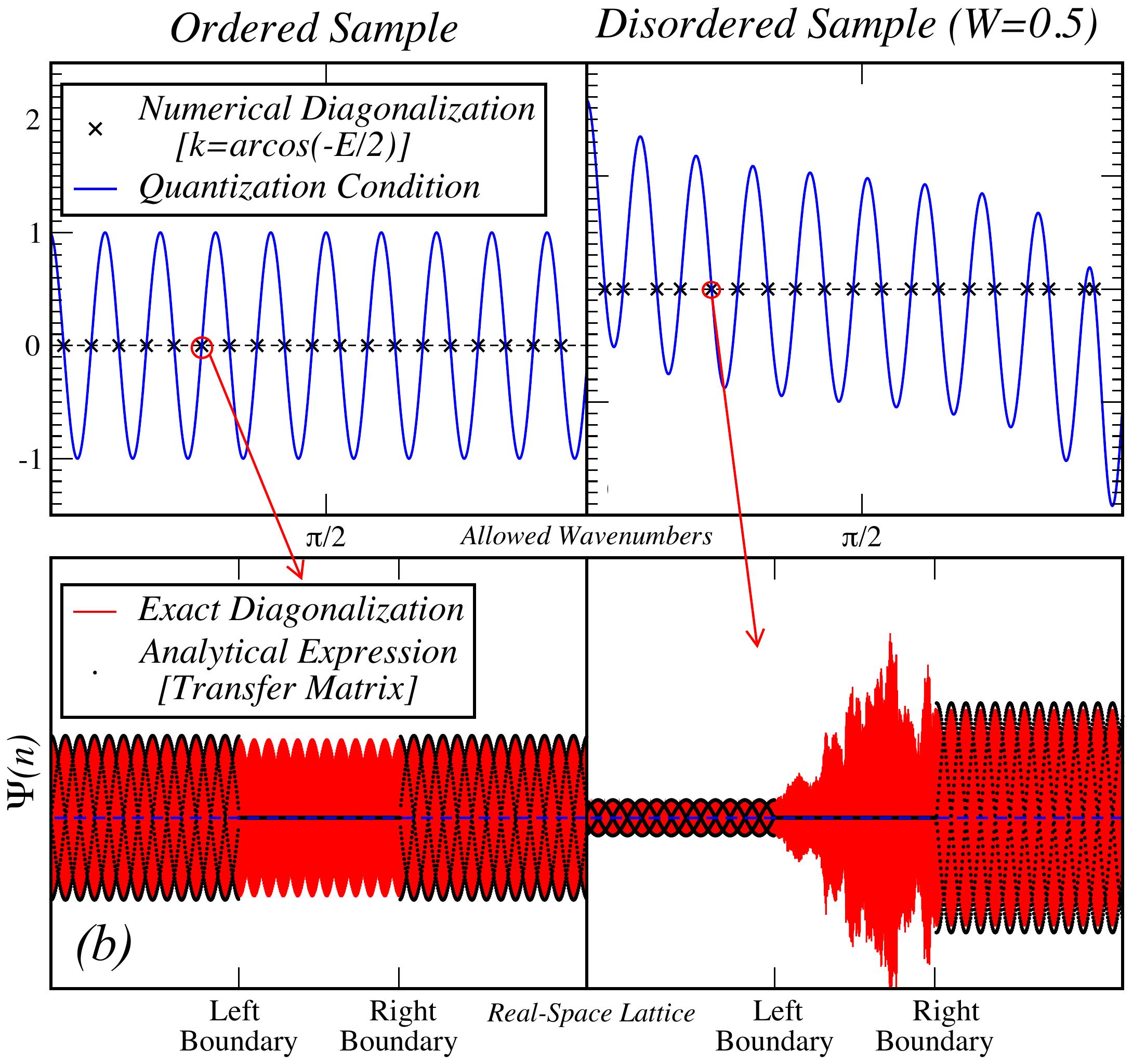}
\par\end{centering}
\caption{\label{fig:ComparisonwithNumericals}\textsl{(a) }Scheme of the procedure
of replacing the central sample by an effective momentum-dependent
transfer matrix, $\mathcal{M}\left(k\right)$. \textsl{(b)} Comparison
between the eigenvalues and eigenstates obtained from the numerical
diagonalization of a system with finite leads of size $L_{l}=8192$
and a sample with $L_{s}=512$ sites, and the ones obtained using
the transfer matrix method. The left panels correspond to a case without
disorder, while the right ones to disordered central sample. The upper
panels compare the wave numbers obtained from the eigenvalues of the
numerical diagonalization with the zeros of the analytical quantization
condition {[}Eq. (\ref{eq:QuantizationCondition}){]}, while the lower
panels compare the corresponding wave functions of one of eigenstates
(signaled by the red arrow).}
\end{figure}

\noindent Combining Eqs.~(\ref{eq:wavefunctions_leads})-(\ref{eq:boundary_condition})
one arrives at the following general expression for the wave functions:

\vspace{-0.5cm}

\begin{equation}
\psi_{k}(n)=\frac{1}{\sqrt{N_{k}}}\begin{cases}
\left|t\left(k\right)\right|\sin\left[k\left(n+L_{l}+1\right)\right] & n<0\\
f_{2}\left(k\right)\sin\left[k\left(n-L_{l}-1\right)\right] & n>0
\end{cases},\label{eq:RealSpaceAmplitudes}
\end{equation}
where $N_{k}$ is a normalization factor, which can be determined
in the limit of large leads by approximating, in the same spirit of
Eq.~(\ref{eq:large_lead:approx}),
\begin{equation}
\sum_{n}\left|\psi_{k}(n)\right|^{2}\simeq\sum_{n\in\text{leads}}\left|\psi_{k}(n)\right|^{2}.
\end{equation}

\noindent This finally leads to

\vspace{-0.5cm}
\begin{equation}
N_{k}\simeq L_{l}f_{1}\left(k\right).\label{eq:normalization}
\end{equation}
The functions $f_{1}(k)$ and $f_{2}\left(k\right)$ are defined as

\vspace{-0.5cm}

\begin{subequations}

\begin{align}
f_{1}\left(k\right) & =1+\abs{r\left(k\right)}\cos\left[2k\left(L_{l}+1\right)+\theta\left(k\right)\right]\\
f_{2}\left(k\right) & =\cos\left[2k\left(L_{l}+1\right)+\phi\left(k\right)\right]\nonumber \\
 & +\abs{r\left(k\right)}\cos\left[\theta\left(k\right)-\phi\left(k\right)\right],
\end{align}
\end{subequations} and where $k$ is constrained to verify the following
quantization condition:

{\small{}
\begin{equation}
\sin\left[2k\left(L_{l}+1\right)+\phi\left(k\right)\right]=\abs{r\left(k\right)}\sin\left[\theta\left(k\right)-\phi\left(k\right)\right].\label{eq:QuantizationCondition}
\end{equation}
}{\small\par}

Notice that the solution of this last condition, together with the
relation $k=\arccos\left(-E/\left(2w\right)\right)$, allows us to
determine the eigenenergies corresponding to delocalized states. In
the lower panels of Fig.~\ref{fig:ComparisonwithNumericals}~(b),
we exemplify the validity of this statement by comparing the wave
functions obtained from the numerical diagonalization of $\mathcal{H}_{0}$,
with $L_{l}=8192$ sites, to the semi-analytical expressions of Eq.~(\ref{eq:RealSpaceAmplitudes}).
The wave numbers obtained from the numerical diagonalization, i.e.
$k=\arccos\left(-E/2\right)$, are also seen to coincide perfectly
with the roots of Eq.~(\ref{eq:QuantizationCondition}) {[}see the
upper panels of Fig.~\ref{fig:ComparisonwithNumericals}\,(b){]}.

Note also that Eqs.~(\ref{eq:RealSpaceAmplitudes}) and (\ref{eq:QuantizationCondition})
reduce to the usual result for the eigenstates of a finite open chain,
when $\abs{r\left(k\right)}=0$ and $\phi\left(k\right)=k\left(L_{l}-1\right)$
is the phase accumulated by a plane-wave crossing the internal bonds
of an ordered sample, i.e.,

\begin{equation}
\psi_{k}(n)=\frac{1}{\sqrt{L_{l}}}\begin{cases}
\sin\left[k\left(n+L_{l}+1\right)\right],\qquad\;\;n<0\\
\left(-1\right)^{p}\sin\left[k\left(n-L_{l}-1\right)\right],n>0
\end{cases}\label{RealSpaceAmplitudes-1}
\end{equation}
with $k=\pi p/\left(L+1\right)$ and $p=1,...,L$. Moreover, these
states are non-degenerate and also alternately symmetrical and antissymmetrical
under parity ($n\to-n$), which just reflects that same symmetry of
the clean Hamiltonian.

With the knowledge of the eigenstates wave functions of the leads,
Eq.~(\ref{eq:RealSpaceAmplitudes}), we can write the matrix elements
$\Gamma_{k,q}$ and $\Pi_{k,q}^{n}$. With the approximation of Eq.~(\ref{eq:large_lead:approx}),
we can evaluate $\Gamma_{k,q}$ analytically, obtaining{\small{}
\begin{multline}
\Gamma_{k,q}\simeq\Delta V\frac{\abs{t\left(k\right)}\abs{t\left(q\right)}-f_{2}\left(k\right)f_{2}\left(q\right)}{8L_{l}\sqrt{f_{1}\left(k\right)f_{1}\left(q\right)}}\times\\
\times\left\{ \frac{\sin\left[\left(q-k\right)\left(L_{l}+\frac{1}{2}\right)\right]}{\sin\left(\frac{q-k}{2}\right)}-\frac{\sin\left[\left(q+k\right)\left(L_{l}+\frac{1}{2}\right)\right]}{\sin\left(\frac{q+k}{2}\right)}\right\} .\label{eq:GammaTransferMatrix}
\end{multline}
}{\small\par}

As for the matrix elements $\Pi_{k,q}^{n}$, from the definition Eq.~(\ref{eq:Gamma_definition}),
for bonds in the left lead ($n<-1$), and after some simple manipulations,
we obtain{\small{}
\begin{multline}
\Pi_{k,q}^{n<-1}=\frac{\left|t\left(k\right)\right|\left|t\left(q\right)\right|}{L_{l}\sqrt{f_{1}\left(k\right)f_{1}\left(q\right)}}\times\\
\times\left\{ \sin\left(\frac{k-q}{2}\right)\sin\left[\left(k+q\right)\left(n+L_{l}+\frac{3}{2}\right)\right]\right.\\
\left.-\sin\left(\frac{k+q}{2}\right)\sin\left[\left(k-q\right)\left(n+L_{l}+\frac{3}{2}\right)\right]\right\} ,\label{eq:PiTransferMatrix}
\end{multline}
}while for the current in bonds of the right lead ($n>1$), we obtain
a similar result after replacing $\left|t\left(k\right)\right|\left|t\left(q\right)\right|\rightarrow f_{2}\left(k\right)f_{2}\left(q\right)$
and $L_{l}\rightarrow-L_{l}-2$ in Eq.\,(\ref{eq:PiTransferMatrix}).

\subsection{Continuum regime of the Kubo formula}

When analyzing the time-dependent Kubo formula of Eq.~(\ref{eq:LRTCurrent-1}),
we must take into account that there are actually two distinct time
scales: \textsl{(1)} the observation time, $t$, and \textsl{(2)}
the scale associated with the spacing between the discrete energy
levels of the finite chain. The latter is proportional to the length
of the leads and, as discussed in the Sec.~\ref{sec:NumericalResults},
is associated to the recurrence time $t_{r}\sim2L_{l}/v_{\text{F}}$.

As expected and confirmed in Sec.~\ref{sec:NumericalResults}, the
quasi-steady-state regime which approximates the Landauer transport
regime of semi-infinite leads, emerges when we take $T,L_{l}\rightarrow\infty$
(with $T=tw/\hbar$ being the time in dimensionless units), but while
keeping $T\ll L_{l}$. In such case, all the transients have died
out, but the system is still far away from getting into the regime
where current inversions occur. Furthermore, Eq.~(\ref{eq:LRTCurrent-1})
includes a factor of $\sin\left(\Delta\varepsilon_{\alpha\beta}t/\hbar\right)/\Delta\varepsilon_{\alpha\beta}$,
which is an emergent $\delta$-function in the limit $t\to+\infty$,
with a broadening of $\hbar t^{-1}$ in energy. This factor actually
acts as a spectral filter which kills-off the contributions coming
from pairs of eigenstates having an energy separation larger than
$\hbar t^{-1}$. Hence, we will show in this section how the approximately
time-independent quasi-steady-state current emerges, when we are in
the limit $T,L_{l}\rightarrow\infty$, with $T\ll L_{l}$, such that
there are many eigenvalues inside the interval $\left[\varepsilon_{\text{F}}-\hbar t^{-1},\varepsilon_{\text{F}}+\hbar t^{-1}\right]$.
We will refer to this limit as the \emph{continuum regime}.

\vspace{-0.5cm}

\subsubsection{Approximate form of $\Gamma_{k,q}$ and $\Pi_{k,q}^{n}$ matrices
in the continuum regime}

We start by noting that, in the continuum limit, since only states
close to the Fermi energy contribute, it suffices to obtain the matrix
elements $\Gamma_{k,q}$ and $\Pi_{k,q}^{n}$ between states where
$k-q$ is small and $k,q\simeq k_{\text{F}}$. In the limit of $k-q\rightarrow0$,
the first term of Eq.~(\ref{eq:GammaTransferMatrix}) dominates over
the second. Therefore, we can approximate it as

\vspace{-0.6cm}
\begin{equation}
\Gamma_{k,q}\simeq\Delta V\frac{\abs{t\left(k\right)}\abs{t\left(q\right)}-f_{2}\left(k\right)f_{2}\left(q\right)}{8L_{l}\sqrt{f_{1}\left(k\right)f_{1}\left(q\right)}}\frac{\sin\left[\left(k-q\right)L_{l}\right]}{\sin\left(\frac{k-q}{2}\right)},\label{eq:Gamma_approx}
\end{equation}
where, in the of limit $L_{l}\rightarrow\infty$ , we approximated
$\sin\left[\left(q-k\right)\left(L_{l}+\frac{1}{2}\right)\right]\simeq\sin\left[\left(q-k\right)L_{l}\right]$.
Doing the same for $\Pi_{k,q}^{n}$, we obtain

\vspace{-0.4cm}
\begin{equation}
\Pi_{k,q}^{n<-1}\simeq-\frac{\abs{t\left(k\right)}\abs{t\left(q\right)}}{L_{l}\sqrt{f_{1}\left(k\right)f_{1}\left(q\right)}}\sin\left(k_{\text{F}}\right)\sin\left[\left(k-q\right)L_{l}\right],\label{eq:Pi_left_approx}
\end{equation}
where we assumed that $\left|n\right|\ll L_{l}$, when approximating
$\sin\left[\left(k-q\right)\left(n+L_{l}+\frac{3}{2}\right)\right]\simeq\sin\left[\left(k-q\right)L_{l}\right]$.
This justifies why in the quasi-steady-state regime, the current is
approximately uniform, if we are away from the chain's extremities.
For the current on the right lead, we obtain a similar result, namely,

\vspace{-0.4cm}
\begin{equation}
\Pi_{k,q}^{n>1}\simeq\frac{f_{2}\left(k\right)f_{2}\left(q\right)}{L_{l}\sqrt{f_{1}\left(k\right)f_{1}\left(q\right)}}\sin\left(k_{\text{F}}\right)\sin\left[\left(k-q\right)L_{l}\right].\label{eq:Pi_right_approx}
\end{equation}

Now, we note that for a chain without any disorder, the matrix elements
of $\Gamma_{k,q}$ will only be non-zero if the states labeled to
$k$ and $q$ have opposite parities. This selection rule stems from
the fact that the fully ordered chain is symmetric under inversion
and therefore its eigenstates will have a well-defined parity. Since
the applied potential $v_{n}^{\text{e}}$ is an odd perturbation,
it only couples states of opposite parities. In the presence of a
general disorder in the central sample, we no longer have inversion
symmetry. Nevertheless, one may still expect that in the limit $L_{l}\gg L_{s}$,
the breaking of the symmetry is small and an approximate selection
rule should emerge. Indeed, this is the case. In order to obtain this
approximate selection rule for a sample with disorder, we notice that
although we can no longer classify the states as even and odd, given
the quantization condition Eq.~(\ref{eq:QuantizationCondition}),
which involves $\sin\left[2k\left(L_{l}+1\right)+\phi\left(k\right)\right]${\small{},
}we can classify the states as $+$ and $-$ according to the sign
of $\cos\left[2k\left(L_{l}+1\right)+\phi\left(k\right)\right]$:{\small{}
\begin{multline}
\cos\left[2k^{\pm}\left(L_{l}+1\right)+\phi\left(k^{\pm}\right)\right]=\\
=\pm\sqrt{1-\abs{r\left(k^{\pm}\right)}^{2}\sin^{2}\left[\theta\left(k^{\pm}\right)-\phi\left(k^{\pm}\right)\right]}.\label{eq:CosClassification}
\end{multline}
}{\small\par}

\begin{figure}[H]
\begin{centering}
\includegraphics[width=8.5cm]{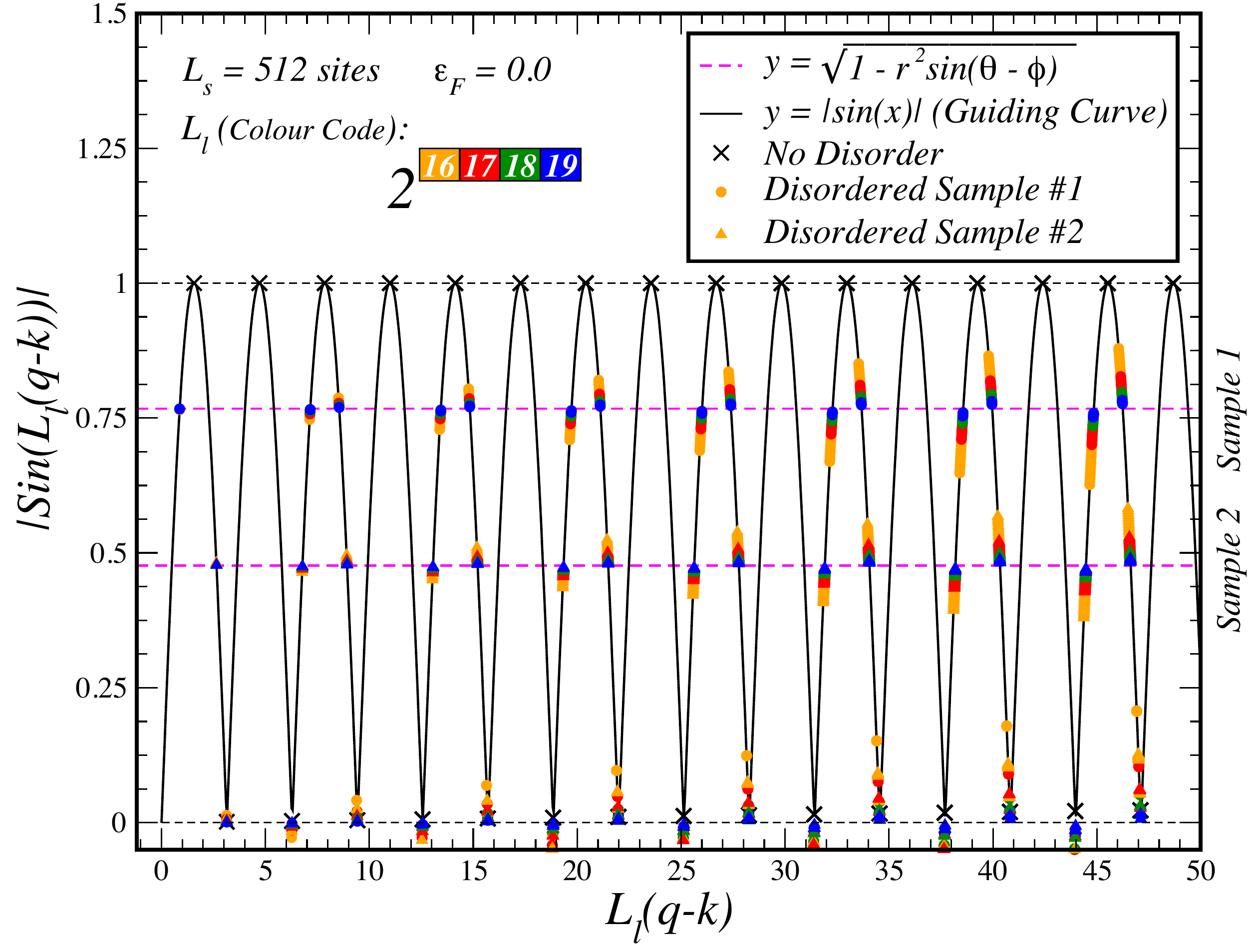}
\par\end{centering}
\caption{\label{fig:SinScaling}Scatter plot of $\protect\abs{\sin\left[\left(k-q\right)L_{l}\right]}$
versus $\left(q-k\right)L_{l}$ for the allowed values of $k,q$ and
different lead sizes ($L_{l}=2^{16}$---$2^{19}$ sites). The different
data sets correspond to four different samples, one ordered ($\times$)
and two disordered ones ({\scriptsize{}$\bigcirc$} and $\triangle$),
which were randomly chosen. The dashed magenta curves correspond to
the asymptotic limits of $k-q\rightarrow0$ and $L_{l}\rightarrow\infty$,
as given by Eq.~(\ref{eq:Effective_SelectionRule}).}
\end{figure}

\noindent For an ordered or symmetrically disordered sample, this
reduces to a labeling of eigenstates as even or odd, respectively,
under a parity transformation, $n\to-n$. With such a classification,
it can be shown (see Appendix~\ref{appx:SelectionRule}) that in
the limits of $k-q\rightarrow0$ and $L_{l}\rightarrow\infty$, one
obtains the following effective selection rule:
\begin{multline}
\lim_{L_{l}\rightarrow\infty}\left|\sin\left[\left(k^{\sigma}-q^{\sigma^{\prime}}\right)L_{l}\right]\right|=\left(1-\delta_{\sigma,\sigma^{\prime}}\right)\times\\
\times\sqrt{1-\abs{r\left(k_{\text{F}}\right)}^{2}\sin^{2}\left(\theta\left(k_{\text{F}}\right)-\phi\left(k_{\text{F}}\right)\right)}.\label{eq:Effective_SelectionRule}
\end{multline}

\begin{figure}[H]
\begin{centering}
\includegraphics[width=8.4cm]{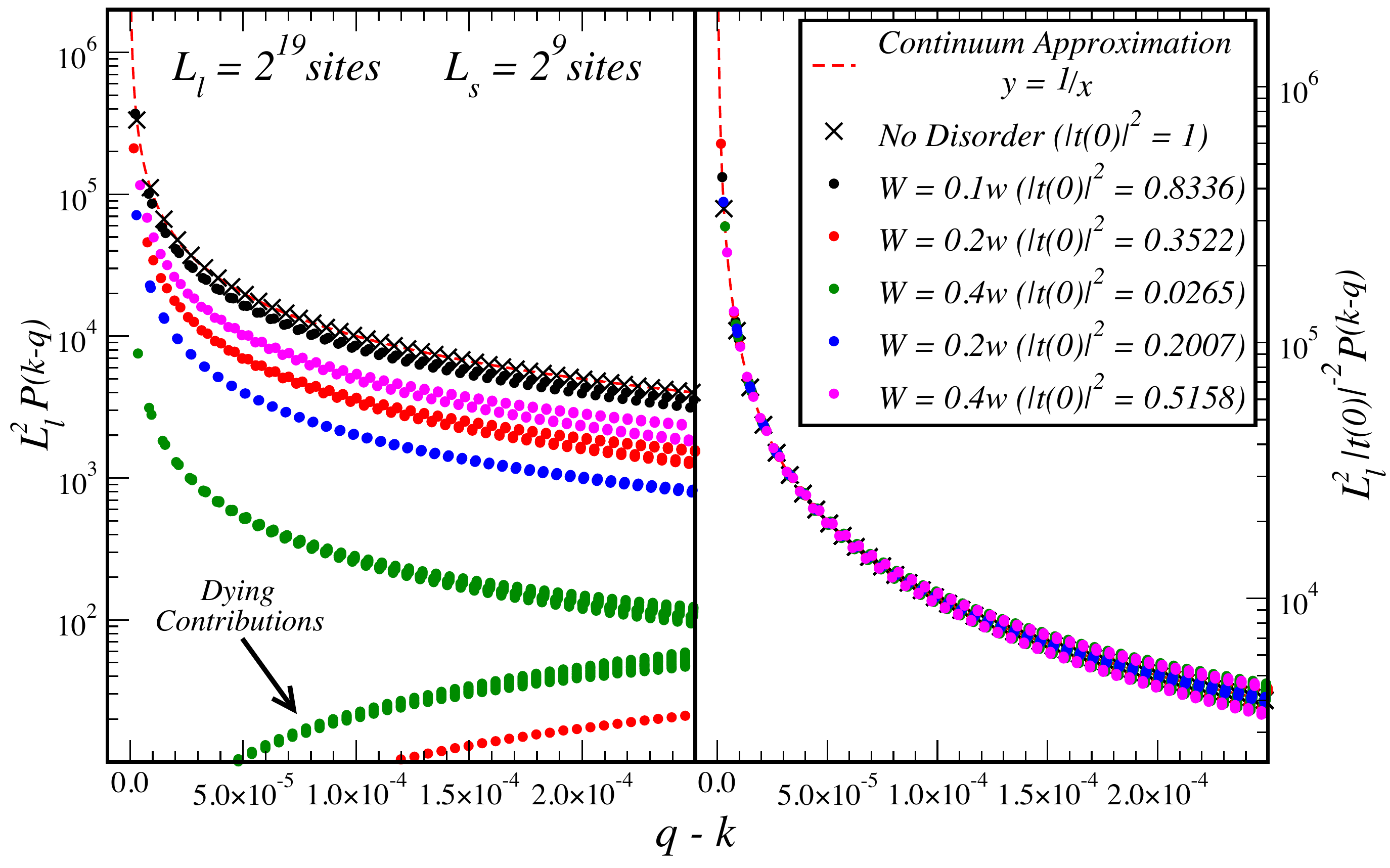}
\par\end{centering}
\begin{centering}
\includegraphics[width=8.4cm]{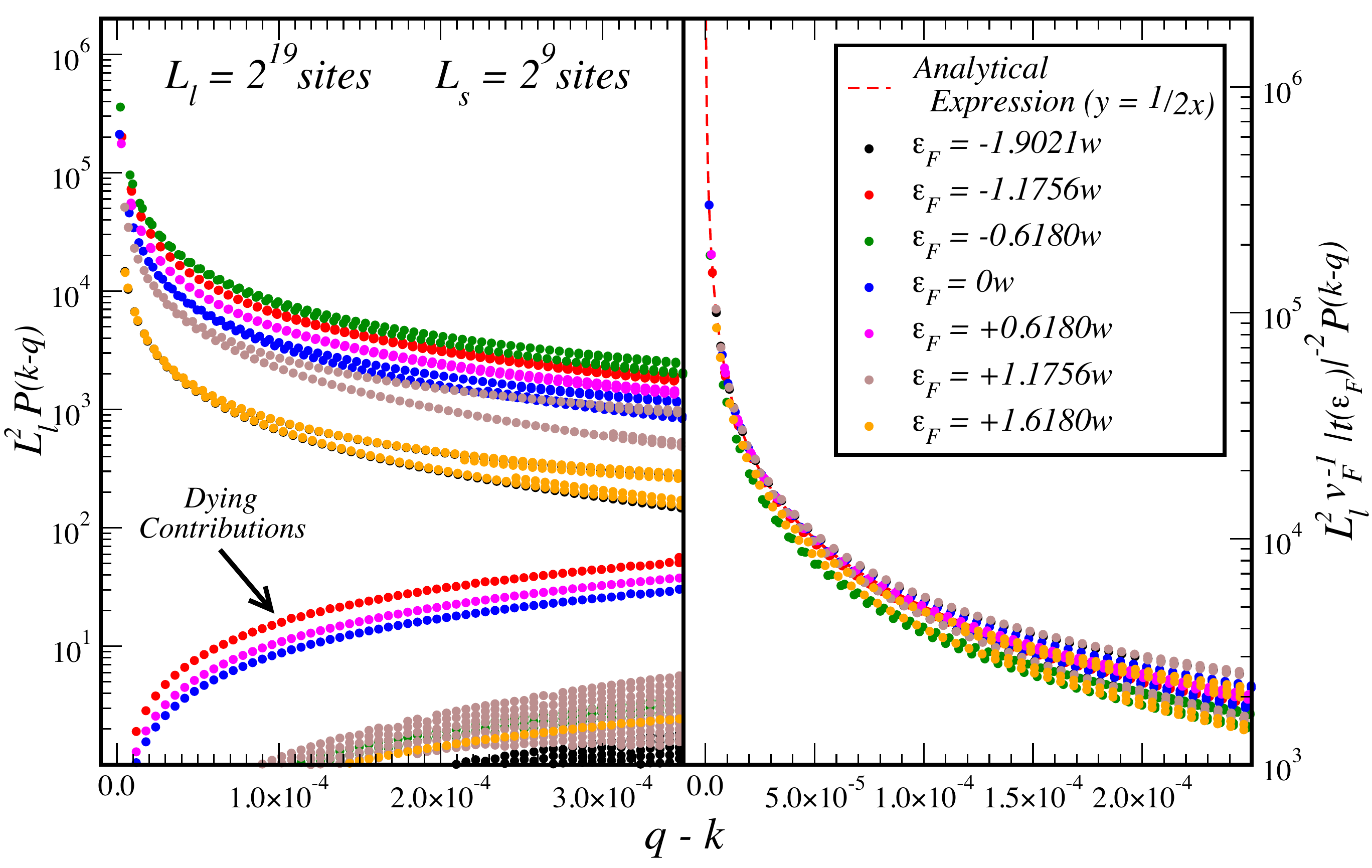}
\par\end{centering}
\caption{\label{FullPrefactors}Plots of the full prefactors $\mathcal{P}\left(\varepsilon_{\text{F}},k-q\right)$
for different central samples at half-filling (upper panels) and different
Fermi energies, $\varepsilon_{\text{F}}$ (lower panels). The collapse
of all the data into the red dashed curves justifies the validity
of expression of Eq. (\ref{eq:FullPrefactorEq}) for states close
to the Fermi level. In both panels, $\mathcal{P}_{k,q}$ is measured
in units of $w^{2}/\hbar$.}
\end{figure}

\noindent with $\sigma,\sigma'=\pm$ and which immediately implies
that $\Gamma_{k,q}\simeq0$, if $k,q$ are in the same class as $L_{l}\rightarrow\infty$.
This is an approximate selection rule, analogous to the one which
exists in the clean case, but which only emerges when $L_{l}\to\infty$.
In Fig.~\ref{fig:SinScaling}, we represent the values of $\left|\sin\left[\left(k-q\right)L_{l}\right]\right|$
as a function of $\left(k-q\right)L_{l}$, for allowed values of $k$
and $q$. We can clearly see that for some data points $\left|\sin\left[\left(k-q\right)L_{l}\right]\right|\rightarrow0$
as $L_{l}$ increases, while other data points tend to a finite value,
which is given by the sample-specific value, $\sqrt{1-\abs{r\left(k_{\text{F}}\right)}^{2}\sin^{2}\left(\theta\left(k_{\text{F}}\right)-\phi\left(k_{\text{F}}\right)\right)}$.\,\footnote{As visible in Fig.\,\ref{fig:SinScaling}, there is small deviation
of the data points obtained for the ordered central sample from the
theoretical value $\abs{\sin\left(L_{l}\left(q^{\sigma}-k^{\sigma'}\right)\right)}=1-\delta_{\sigma,\sigma'}$.
These results seem incompatible with an exact parity selection rule
for $\Gamma_{k,q}$, for this case, however, they are not. This artifact
is due to the fact that the data shown was calculated from the numerical
diagonalization of $\mathcal{H}_{0}$ and then $k=\arccos\left(-E/2\right)$
was used to obtain the respective wave numbers. This procedure takes
into account the finite dimension of the central sample and hence
the allowed wave numbers are of the form $k^{\sigma}=\pi n/\left(2L_{l}+L_{s}+1\right)$,
with $\sigma=\left(-1\right)^{n}$. Since the expression of $\Gamma_{k,q}$
which is proportional to $\sin\left(L_{l}\left(q-k\right)\right)$
is only valid under an approximation which ignores $L_{s}$ {[}i.e.
Eq.\,(\ref{eq:large_lead:approx}){]}, the parity selection rule
appears to be approximate as well. At any rate, it may be proven,
by symmetry, that this rule is actually true for any value of $L_{l}$,
if one considers the full expression for $\Gamma_{k,q}$.}

We point out that in the case of symmetric disorder profile, one can
derive from the properties of the transfer matrix that $\phi\left(k\right)-\theta\left(k\right)=\pm\pi/2$.
In this case, one immediately sees that the scattering wave functions
of Eq. (\ref{eq:RealSpaceAmplitudes}) reduce to the same form as
in Eq.~(\ref{RealSpaceAmplitudes-1}), with the parity determined
by the class to which it belongs. In such a case, the $\abs{t\left(k\right)}$
factor of the $\Gamma$ and $\Pi$ matrices comes only from this effect,
since the functions $f_{1}\left(k\right)$ and $f_{2}\left(k\right)$
are exactly the same as in the non-disordered case. Just the allowed
$k$'s are different.

Having established this effective selection rule, we can expand the
prefactors of Eqs.~(\ref{eq:Gamma_approx})---(\ref{eq:Pi_right_approx})
around $k_{\text{F}}$. Taking into account that the only significant
contributions come from pairs of states belonging to different classes,
we may use Eqs.~(\ref{eq:QuantizationCondition}) and (\ref{eq:CosClassification})
to write, for $k,q\simeq k_{\text{F}}$,
\begin{flushleft}
\vspace{-0.7cm}
\begin{subequations}
\begin{multline}
f_{1}\left(k\right)f_{1}\left(q\right)\simeq\\
\simeq\left(1-\abs{r\left(k_{\text{F}}\right)}^{2}\sin^{2}\left[\theta\left(k_{\text{F}}\right)-\phi\left(k_{\text{F}}\right)\right]\right)\left|t\left(k_{\text{F}}\right)\right|^{2},
\end{multline}
\begin{equation}
f_{2}\left(k\right)f_{2}\left(q\right)\simeq-\left|t\left(k_{\text{F}}\right)\right|^{2}.
\end{equation}
\end{subequations}Using these approximations in Eq.~(\ref{eq:Gamma_approx}),
we obtain
\begin{align}
\Gamma_{k,q} & \simeq\Delta V\frac{\left|t\left(k_{\text{F}}\right)\right|}{4L_{l}}\frac{1}{\sin\left(\frac{k-q}{2}\right)}\nonumber \\
 & \simeq\Delta V\frac{\left|t\left(k_{\text{F}}\right)\right|}{2L_{l}}\frac{1}{k-q},\label{eq:Gamma_approx_final}
\end{align}
and from Eqs.~(\ref{eq:Pi_left_approx}) and (\ref{eq:Pi_right_approx})
we obtain
\begin{equation}
\Pi_{k,q}^{n<-1}\simeq\Pi_{k,q}^{n>1}\simeq-\frac{\left|t\left(k_{\text{F}}\right)\right|}{L_{l}}\sin\left(k_{\text{F}}\right).\label{eq:Pi_approx_final}
\end{equation}
In the following, we will use Eqs.\,(\ref{eq:Gamma_approx_final})
and (\ref{eq:Pi_approx_final}) to obtain the Landauer current from
the Kubo formula of Eq.\,(\ref{eq:LRTCurrent-1})
\par\end{flushleft}

\subsubsection{Continuum limit expression for the stationary current: Emergence
of Landauer transport}

Using Eqs.~(\ref{eq:Gamma_approx_final}) and (\ref{eq:Pi_approx_final}),
we can write the time-dependent Kubo formula Eq.~(\ref{eq:LRTCurrent-1})
as
\begin{equation}
I^{n}\left(t\right)=\frac{e}{2}^{2}\sideset{}{^{\prime}}\sum_{k,q}\mathcal{P}_{k,q}\frac{\sin\left(\Delta\varepsilon_{k,q}t/\hbar\right)}{\Delta\varepsilon_{k,q}},\label{eq:emergent_Landauer}
\end{equation}
where the primed sum in Eq.~(\ref{eq:emergent_Landauer}) means that
only pairs of states $\left(k,q\right)$ of opposite classes are included
in the sum, due to the emergent selection rule. We also introduced
the quantity $\mathcal{P}_{k,q}$, which is defined as
\begin{align}
\mathcal{P}_{k,q} & =\frac{2w}{\hbar}\Pi_{k,q}^{n}\Gamma_{k,q}\Delta f_{k,q}\nonumber \\
 & \simeq-\frac{1}{2L_{l}^{2}}\frac{\left|t\left(k_{\text{F}}\right)\right|^{2}v_{\text{F}}^{2}\hbar}{\abs{\Delta\varepsilon_{k,q}}}\Delta V,\label{eq:FullPrefactorEq}
\end{align}
where we approximated $v_{\text{F}}\hbar\left(k-q\right)\simeq\Delta\varepsilon_{k,q}$,
with $v_{\text{F}}=2w\sin\left(k_{\text{F}}\right)/\hbar$. To make
Eqs.\,(\ref{eq:emergent_Landauer}) and (\ref{eq:FullPrefactorEq})
more clear, we remark that this definition of the current is no longer
dependent on the condition $k<k_{\text{F}}<q$, and $\mathcal{P}_{k,q}$
is actually symmetrical upon exchange of the indices. The above equation
also shows that, for $k,q\simeq k_{\text{F}}$, the latter approximately
only a function of the difference in eigenenergies. This result was
checked numerically as seen in Fig.~\ref{FullPrefactors}, where
we can see that for a wide variety of disordered samples and different
values of the Fermi energy, all the values of $\mathcal{P}_{k,q}$
{[}calculated directly from the wave functions in Eq.~(\ref{RealSpaceAmplitudes-1}){]}
fall into the curve Eq.\,(\ref{eq:FullPrefactorEq}).

The time-dependent current in the continuum regime can thus be written
as

\vspace{-0.3cm}

\begin{multline}
I^{n}\left(t\right)=\frac{e^{2}}{\hbar}\left|t\left(k_{\text{F}}\right)\right|^{2}\left(v_{\text{F}}\hbar\right)^{2}\Delta V\\
\times\int_{-\infty}^{+\infty}d\left(\Delta\varepsilon\right)\frac{\sin\left(\Delta\varepsilon t/\hbar\right)}{\Delta\varepsilon\abs{\Delta\varepsilon}}\varrho\left(\Delta\varepsilon\right),\label{eq:current_JDoCS}
\end{multline}
where we introduced the \emph{joint density of contributing states}
(JDoCS), $\varrho$, as

\vspace{-0.5cm}

\begin{equation}
\varrho\left(\epsilon_{\text{F}},\Delta\varepsilon\right)=\frac{1}{4L_{l}^{2}}\sideset{}{^{'}}\sum_{\overset{k,q}{}}\delta\left(\Delta\varepsilon-\Delta\varepsilon_{k,q}\right).\label{eq:JDoCS}
\end{equation}

The restricted summation in Eq.\,(\ref{eq:JDoCS}) already takes
into account the emergent selection rule of Eq.~(\ref{eq:Effective_SelectionRule}).
In Appendix~\ref{appx:CalculationJDoCS}, we show that this quantity,
in the limit $L_{l}\to\infty$, can be written in terms of the density
of states of each class in a fully clean system and its expression
for small enough $\abs{\Delta\varepsilon}$ is simply

\vspace{-0.3cm}

\begin{equation}
\lim_{L_{l}\to\infty}\left[\varrho\left(\varepsilon_{\text{F}},\Delta\varepsilon\right)\right]=\frac{\abs{\Delta\varepsilon}}{2\pi^{2}\left(4w^{2}-\varepsilon_{\text{F}}^{2}\right)}+\mathcal{O}\left[\Delta\varepsilon^{2}\right].\label{LinearizedDPEP}
\end{equation}
Hence, when Eq.~(\ref{LinearizedDPEP}) is plugged into Eq.~(\ref{eq:current_JDoCS}),
we get

\begin{multline}
I^{n}\left(t\right)=\frac{e^{2}}{2\pi^{2}\hbar}\frac{\left|t\left(k_{\text{F}}\right)\right|^{2}\left(v_{\text{F}}\hbar\right)^{2}\Delta V}{4w^{2}-\varepsilon_{\text{F}}^{2}}\\
\times\int_{-\infty}^{\infty}d\left(\Delta\varepsilon\right)\frac{\sin\left(\Delta\varepsilon t/\hbar\right)}{\Delta\varepsilon}.\label{eq:current_JDoCS-1}
\end{multline}
Finally, Eq.\,(\ref{eq:current_JDoCS-1}) together with the facts
that

\vspace{-0.6cm}

\begin{equation}
\lim_{T\to\infty}\left[\frac{\sin\left[xT\right]}{x}\right]=\pi\delta\left(x\right),
\end{equation}
and $v_{F}\hbar=\sqrt{4w^{2}-\varepsilon_{F}^{2}}$, yields a steady-state
current
\begin{equation}
I^{n}\left(t\right)=\frac{e^{2}}{h}\left|t\left(k_{F}\right)\right|^{2}\Delta V,\label{eq:Landauer_Current}
\end{equation}
which is precisely the linear Landauer steady-state current for a
two-terminal one-dimensional device.

Notice, that in the derivation of this result from the time-dependent
Kubo formula, it is essential that $t,L_{l}\rightarrow\infty$ with
$wt/\hbar\ll L_{l}$, such that the $\varrho\left(\Delta\varepsilon\right)$
can be evaluated in the limit of $L_{l}\rightarrow\infty$, while
the factor $\sin\left(\Delta\varepsilon t/\hbar\right)/\Delta\varepsilon$
is treated as as emergent $\delta$-function. When $wt/\hbar\apprge L_{l}$,
then there will be few pairs of states with $\Delta\varepsilon_{k,q}\in\left[\varepsilon_{F}-\hbar t^{-1},\varepsilon_{F}+\hbar t^{-1}\right]$,
and we can no longer treat $\sin\left(\Delta\varepsilon t/\hbar\right)/\Delta\varepsilon$
as a $\delta$-function. When this happens, we start observing recurrences
in the current as reported in Sec.~(\ref{sec:NumericalResults}).

\vspace{-0.4cm}

\section{\label{sec:Conclusions}Conclusions}

In this work, we investigated how a quasi-steady-state particle transport
regime emerges across disordered samples coupled to large, but finite
leads which are subjected to a potential bias. In order to do so,
we have studied time-dependent transport, both numerically and semi-analytically,
in a non-interacting and one-dimensional tight-binding chain, with
open boundary conditions, where the central region is an extended
disordered sample, and the rest of the chain acts as a pair of finite,
but otherwise perfect leads.

For large lead sizes, and sufficiently large bias, a quasi-steady-state
regime emerges at intermediate times, after the transient behavior
has died out and before inversions in the current are observed. The
current in the quasi-steady-state is approximately constant in time
and homogeneous in space (if measured at points far away from the
chain's extremities). Furthermore, the value of the current in the
quasi-steady-state coincides with the one predicted by the Landauer
formula for semi-infinite leads, independently of the initial condition
of the system (\textit{partitioned} or \textit{partition-free}). These
results amount to an exemplification and extension to finite systems
of the results of Stefanucci \textit{et al}\,\citep{stefanucci_time-dependent_2004}
on the establishment of a steady-state regime of transport in samples
which are attached to infinite leads.

We have found that the quasisteady state is established, for both
initial conditions, after a stabilization time $t_{\text{stab}}\approx2L_{\text{s}}/v_{\text{F}}$.
Physically, this can be interpreted as the time taken by a Fermi-level
state to probe the disordered landscape inside the central sample.
The quasi-steady-state lasts until a \textit{recurrence time} $t_{\text{r}}\approx2L_{\text{l}}/v_{\text{F}}$,
where current inversions start happening. Aside from being related
to the inverse spacing of the energy levels in the system\,\citep{pal_emergence_2018},
this recurrence time may also be interpreted as the time taken by
a Fermi-level electron to leave the sample and return to it, by traveling
back and forth inside a lead. This conclusion was seen to be independent
of the central sample's features, as long as the leads are much larger
than it and transport is ballistic across the disordered sample.

During the quasi-steady-state, persistent finite-size effects are
observed in the\textit{ partition-free} approach as superposed oscillations,
with a period that is inversely proportional to the bias $\Delta V$
and an amplitude which scales to zero as $L_{\text{l}}\to\infty$
but becomes more relevant (relative to $I_{\text{Landauer}}$) for
very small values of $\Delta V$. This effect prevents the onset of
a quasi-steady-state regime for systems prepared in the \textit{partition-free}
setup, if the leads are too small. In the \textit{partitioned }case,
the amplitude of the oscillations superposed on the quasisteady-state
plateaus is not influenced by the size of leads, but instead is damped
as the observation time increases (while keeping $t<t_{\text{r}}$).
Similarly to the \textit{partition-free} case, the amplitude of the
fluctuations increases for smaller biases. These observations seem
to indicate that the observation of a clear quasisteady state requires
some kind of mechanism which scatters the electron's momenta \citep{bushong_approach_2005}.
Here it is provided by the applied potential ramp in the sample, which
becomes a less effective mechanism as $\Delta V\to0$. In both cases,
these finite-size oscillations can be made arbitrarily small if $L_{l}$
is large enough.

In order to shine light on the numerical results, a time-dependent
Kubo formula for the current in the \textit{partition-free} approach,
which is suitable for semi-analytical treatment, was developed for
describing the local time-dependent current due to a small applied
bias. From this formula, it was possible to see that an approximately
time-independent and spatially uniform current emerges in the limit
of large system's size and observation times, $L_{l},t\rightarrow\infty$,
provided $t\ll v_{\text{F}}^{-1}L_{l}$ (in agreement with the recurrence
times observed numerically). These conditions are necessary to treat
the leads as being effectively infinite, in what respects DC transport.
After expressing the eigenfunctions of the disordered central sample
in terms of complex reflection and transmission coefficients, all
the matrix elements appearing in the Kubo formula were evaluated semi-analytically.
The quasi-steady-state current thus obtained was shown to reproduce
the linearized Landauer formula for the current in a two-terminal
device.

We hope that these theoretical predictions of the time scales over
which the quasisteady state occurs and the nature of the finite-size
oscillations can be experimentally tested and guide future research
on mesoscopic transport in fermionic ultra-cold atomic gases in optical
lattices.

\vspace{-0.3cm}

\section{Acknowledgments}

J.M.V.P.L. and J.P.S.P. acknowledge financing of Fundação da Ciência
e Tecnologia, of COMPETE 2020 program in FEDER component (European
Union), through projects POCI-01-0145-FEDER-028887 and UID/FIS/04650/2013.
J.P.S.P. is supported by the MAP-fis PhD grant PD/BD/142774/2018 of
Fundação da Ciência e Tecnologia. B.A. acknowledges financial support
from Fundação para a Ciência e a Tecnologia, Portugal, through Project
Nº CEECIND/02936/2017. Additionally, J.P.S.P. also acknowledges the
hospitality of the University of Central Florida, where part of this
work was done, as well as Dr. Eduardo Mucciolo and Dr. Caio H. Lewenkopf
for the careful reading of the manuscript and suggestions. We also
thank S. M. João, Dr. Nuno M. R. Peres, Dr. Aires Ferreira and Dr.
João M. B. Lopes dos Santos for enlightening discussions about this
work. J.P.S.P. also acknowledges the hospitality of Katherine Vasquez
during his stay in Orlando, Florida. Finally, the authors acknowledge
the comments of the anonymous referees, which proved very helpful
in the improvement of the original manuscript.

\appendix

\section{\label{appx:TimeEvolutionMethod}Review of the recursive Chebyshev
method for quantum time-evolution}

In this appendix, we wish to describe shortly the algorithm used to
time-evolve an arbitrary single-particle state with the full Hamiltonian.
As referred in the main text, the Hamiltonian generating the time-evolution
for positive times, $\mathcal{H}\left(t>0\right)$, is time-independent
and, consequently, the time-evolution operator $\mathcal{U}_{t}$
reads

\vspace{-0.5cm}

\begin{equation}
\mathcal{U}_{t}=e^{-i\mathcal{H}\left(t>0\right)t/\hbar}.
\end{equation}

The method used to calculate $\mathcal{U}_{t}$ for our systems is
based on its exact expansion as a series of Chebyshev polynomials
in $\mathcal{H}\left(t>0\right)$, due to Tal-Ezer\textit{ et al}.
\citep{KPM_Timeevolution_Talezer1984}. Namely, one has

\vspace{-0.6cm}

\begin{equation}
\mathcal{U}_{t}=\sum_{n=0}^{\infty}\frac{2}{1+\delta_{n,0}}\left(-i\right)^{n}J_{n}(\lambda t)T_{n}(\tilde{\mathcal{H}}),\label{ChebyshevExpansion_of_Ut}
\end{equation}
where $\mathcal{\tilde{H}}=\left(\nicefrac{1}{\lambda w}\right)\mathcal{H}$
is a dimensionless Hamiltonian, rescaled by a real parameter $\lambda$
which guarantees that its spectrum is contained inside the interval
$\left]-1,1\right[$, $T_{n}$ is the $n^{\text{th}}$-order Chebyshev
polynomial of the first-kind, $J_{n}(y)$ is a Bessel function of
the first kind and $t$ is a time measured in units of $\frac{\hbar}{w}$.
The key to the method is to avoid the numerical diagonalization of
$\mathcal{H}\left(t>0\right)$, and instead use the recursion relation
for the Chebyshev polynomials,

\vspace{-0.7cm}

\begin{equation}
T_{n+1}(x)=xT_{n}(x)-T_{n-1}(x),\label{ChebyshevRecursion}
\end{equation}
in order to evaluate all the needed $T_{n}\left(\tilde{\mathcal{H}}\right)$,
recursively. For a generic review on the application of Chebyshev
spectral method to physical problems see Ref.\,\citep{weise_kernel_2006}
and references within.

\begin{figure}[H]
\begin{centering}
\includegraphics[scale=0.34]{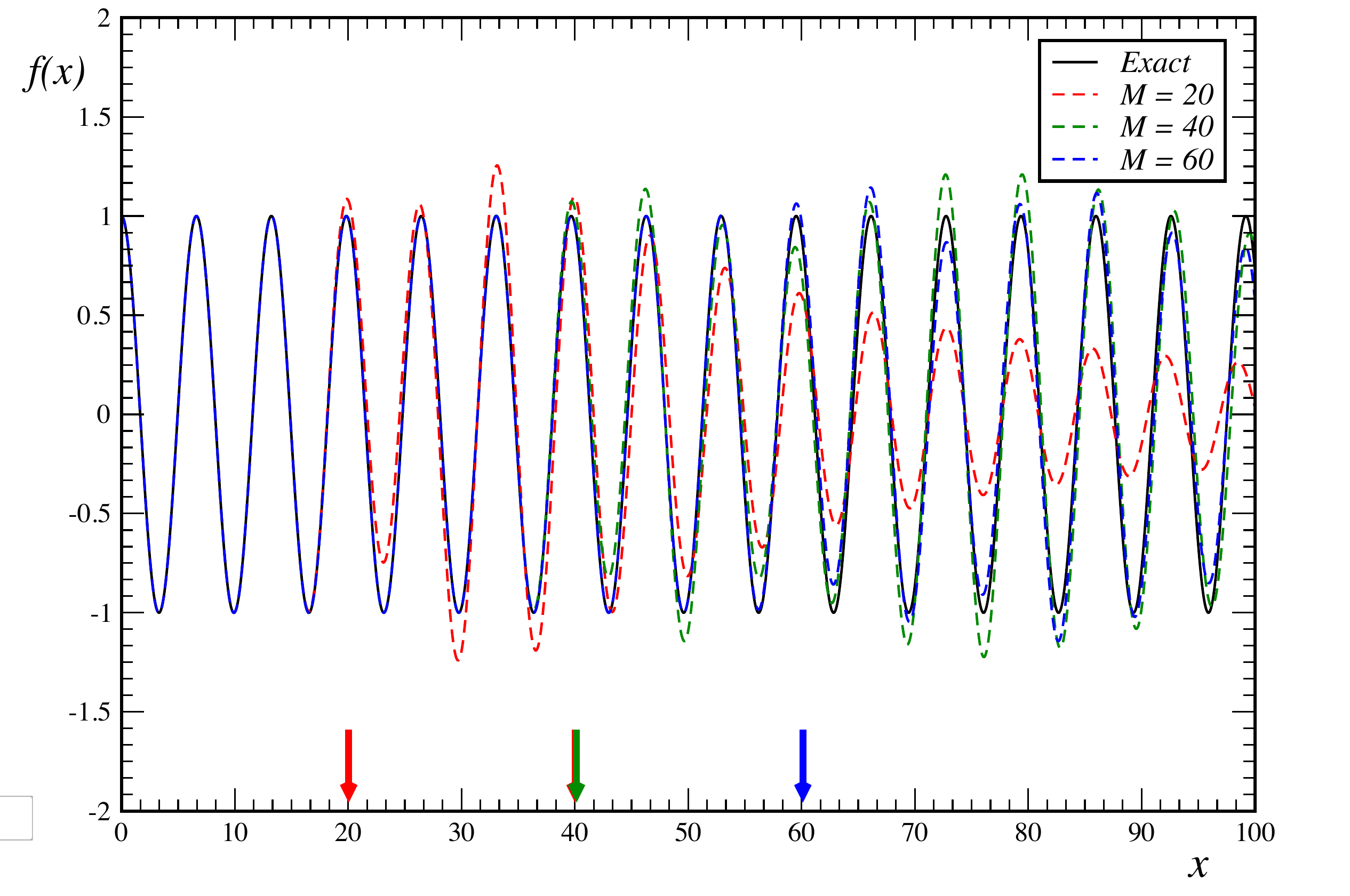}
\par\end{centering}
\caption{\label{ChebyshevConvergence}Comparison between the exact graph for
$f(x)=\text{Re}\left[e^{iyx}\right]$ and successive truncated Chebyshev
series with the first $M=20,40$ and $60$ polynomials. The colored
arrows stand on the values for which the corresponding approximations
starts to fail. The imaginary part has an analogous behavior. (color
online)}
\end{figure}

Furthermore, the Chebyshev series of Eq.\,(\ref{ChebyshevExpansion_of_Ut})
is known to converge rather quickly, meaning that a truncated summation
with $M$ terms is usually enough to describe correctly $\mathcal{U}_{t}$,
provided $M>t\lambda$. This convergence is illustrated in Fig.\,\ref{ChebyshevConvergence}
and in all our calculations, we used $M=8t\lambda$.

Notice that, in order to evaluate the current, we only require to
time-evolve a given single-particle state $\ket{\Psi}$. Therefore,
we do need the full matrix form of $\mathcal{U}_{t}$, but instead
how it acts on an arbitrary state $\ket{\Psi}$. From the expansion
of Eq.\,(\ref{ChebyshevExpansion_of_Ut}), we know that action to
be

\vspace{-0.5cm}

\begin{equation}
\ket{\Psi^{M}(t)}=\sum_{n=0}^{M}\frac{2}{1+\delta_{n,0}}\left(-i\right)^{n}J_{n}(\lambda t)\ket{\Psi_{n}},\label{ChebyshevExpansion_of_state-1}
\end{equation}
where $\ket{\Psi_{n}}=T_{n}\left(\tilde{\mathcal{H}}\right)\ket{\Psi}$
and $M$ is the truncation order of the Chebyshev expansion.\,\footnote{As this is needed in the main text, we remark that for backward time
evolutions, one may simply use the fact that $J_{n}\left(-x\right)=\left(-1\right)^{n}J_{n}\left(x\right)$.} Finally, the first two $\ket{\Psi_{n}}$ can be directly calculated
by the simple forms of $T_{0}\left(x\right)$ and $T_{1}\left(x\right)$,
i.e.

\vspace{-0.5cm}

\begin{subequations}

\begin{align}
\ket{\Psi_{0}} & =T_{0}(\tilde{\mathcal{H}})\ket{\Psi}=\ket{\Psi}\\
\ket{\Psi_{1}} & =T_{1}(\tilde{\mathcal{H}})\ket{\Psi}=\tilde{\mathcal{H}}\ket{\Psi},
\end{align}
\end{subequations}and then the remaining are efficiently calculated
by using the operator generalization of the Chebyshev recursion {[}Eq.\,(\ref{ChebyshevRecursion}){]},
i.e.

\vspace{-0.5cm}

\begin{equation}
\ket{\Psi_{n+1}}=\tilde{\mathcal{H}}\ket{\Psi_{n}}-\ket{\Psi_{n-1}}.
\end{equation}

\section{\label{appx:RecursiveTransferMatrixAppendix}Review of the recursive
transfer matrix method}

In this appendix, we explore a very simple algorithm which allows
us to calculate the transfer matrix $\mathcal{M}\left(k\right)$ of
any given disordered sample, when it is connected to semi-infinite
leads. This method is the same used in the early papers of Andereck
\emph{et al} \citep{andereck_numerical_1980} and Pichard \citep{pichard_one-dimensional_1986}
and allows for the calculation of $\mathcal{M}\left(k\right)$ with
an $\sim\mathcal{O}\left(L_{S}\right)$ number of operations.

For these purposes, it is more useful to re-express the Hamiltonian
of the central sample in a first-quantization language, i.e.,

\vspace{-0.5cm}
\begin{align}
\mathcal{H}_{s} & =\sum_{n=1}^{L_{s}}\varepsilon_{n}\left|n\right\rangle \left\langle n\right|\\
 & \qquad\qquad-\sum_{n=1}^{L_{s}-1}\left(\left|n\right\rangle \left\langle n+1\right|+\left|n+1\right\rangle \left\langle n\right|\right),\nonumber 
\end{align}
where $\left|n\right\rangle $ are the Wannier states of the chain
and $\varepsilon_{n}$ is an on-site energy (in units of the hopping
$w$). To model the connection between the finite sample to the semi-infinite
leads, one has also the following boundary hopping Hamiltonian:

\vspace{-0.4cm}

{\small{}
\begin{equation}
\mathcal{H}_{s}=-\ket 0\bra 1-\ket 1\bra 0-\ket{L_{s}}\bra{L_{s}+1}-\ket{L_{s}+1}\bra{L_{s}}.
\end{equation}
}{\small\par}

\begin{figure}[H]
\begin{centering}
\includegraphics[scale=0.45]{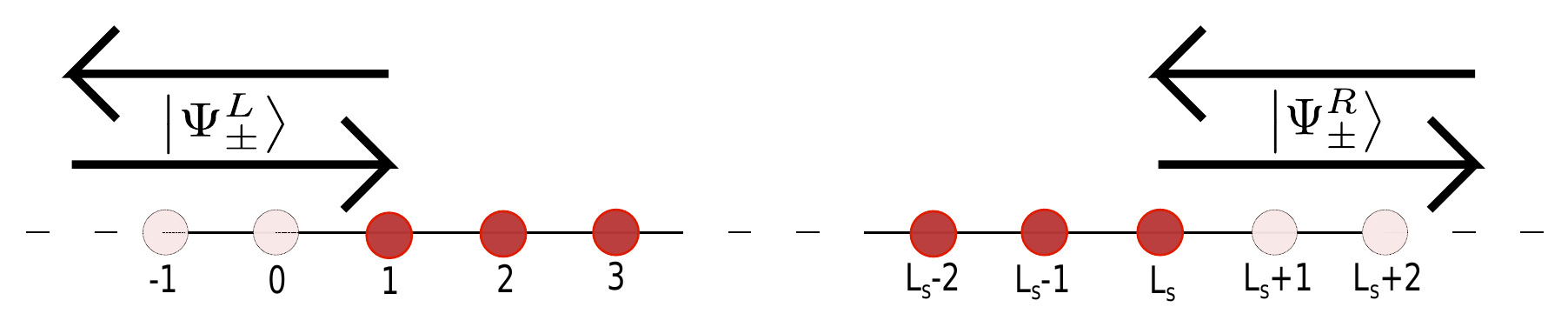}
\par\end{centering}
\caption{\label{fig:TransferMatrixSetup}Schematic representation of the setup
used in the implementation of the transfer matrix method. Red dots
represent the disordered scattering region. The leads are represented
as the lighter red ``ghost'' sites on both sides.}
\end{figure}

The main purpose of this method is to find the scattering states associated
to a particular disorder realization. For that, one must fix the leads'
propagating states, $\ket{\Psi_{\pm}^{L}}$ and $\ket{\Psi_{\pm}^{R}}$,
as the left and right boundary conditions for the problem. This setup
is represented in Fig. \ref{fig:TransferMatrixSetup}, with the counter-propagating
plane waves in the leads being represented as arrows.

\vspace{-0.3cm}

\subsection{Hamiltonian in Real-Space and Boundary Conditions}

The first step towards the definition of the present method is expanding
a scattering state (with wavenumber $k$) in the basis of Wannier
wave functions, i.e.

\vspace{-0.4cm}

\begin{equation}
\ket{\Psi_{k}}=\sum_{n}\psi_{n}\ket n,
\end{equation}
and finally rewriting the time-independent Schrödinger equation, $\mathcal{H}\ket{\Psi_{k}}=E_{k}\ket{\Psi_{k}}$,
in terms of the real-space amplitudes $\psi_{n}$,

\vspace{-0.4cm}

\begin{equation}
E_{k}\psi_{n}=\varepsilon_{n}\psi_{n}-\psi_{n-1}-\psi_{n+1},\label{SchroedingerEq_RealSpace}
\end{equation}
where, by definition, $\varepsilon_{n}=0$ outside of the sample.

As shown in Fig. \ref{fig:TransferMatrixSetup}, the boundary conditions
are to be set as the plane waves defined in Eq. (\ref{eq:wavefunctions_leads})
of the main text. Reminding, one has

\vspace{-0.4cm}

{\small{}
\begin{equation}
\ket{\Psi_{k}^{L}}=\sum_{n=-L_{l}}^{-1}\left[\Psi_{+}^{L}e^{ik\left(n-1\right)}\ket n+\Psi_{-}^{L}e^{-ik\left(n-1\right)}\ket n\right],
\end{equation}
}{\small\par}

\vspace{-0.4cm}

{\small{}
\begin{equation}
\ket{\Psi_{k}^{R}}=\sum_{n=1}^{L_{l}}\left[\Psi_{+}^{R}e^{ik\left(n-L_{s}\right)}\ket n+\Psi_{-}^{R}e^{-ik\left(n-L_{s}\right)}\ket n\right].
\end{equation}
}These states immediately set the amplitudes on the ``ghost'' sites
of Fig.\,\ref{fig:TransferMatrixSetup} to the following values:

\vspace{-0.3cm}

\begin{equation}
\begin{aligned}\psi_{-1} & =\Psi_{+}^{L}e^{-2ik}+\Psi_{-}^{L}e^{2ik},\\
\psi_{0} & =\Psi_{+}^{L}e^{-ik}+\Psi_{-}^{L}e^{ik},\\
\psi_{L_{S}+1} & =\Psi_{+}^{R}e^{ik}+\Psi_{-}^{R}e^{-ik},\\
\psi_{L_{S}+2} & =\Psi_{+}^{R}e^{2ik}+\Psi_{-}^{R}e^{-2ik}.
\end{aligned}
\label{BoundaryConditions}
\end{equation}

\vspace{-0.5cm}

\subsection{Review of the transfer matrix recursive method}

Despite not having the look of a linear algebra problem, Eq.\,(\ref{SchroedingerEq_RealSpace})
may be turned into a matrix recursion equation, when supplemented
by the trivial condition

\vspace{-0.5cm}

\[
\psi_{n}=\psi_{n}.
\]
Hence, we have

\vspace{-0.5cm}

\begin{equation}
\left(\begin{array}{c}
\psi_{n+1}\\
\psi_{n}
\end{array}\right)=\underset{\mathbb{T}_{n}\left(k\right)}{\underbrace{\left(\begin{array}{cc}
\varepsilon_{n}-E_{k} & -1\\
1 & 0
\end{array}\right)}}\cdot\left(\begin{array}{c}
\psi_{n}\\
\psi_{n-1}
\end{array}\right).\label{TransferMatrixOneStep}
\end{equation}
If we now iterate Eq.\,(\ref{TransferMatrixOneStep}), we get the
following relation

\vspace{-0.5cm}

\begin{multline}
\left(\begin{array}{c}
\psi_{L_{s}+2}\\
\psi_{L_{s}+1}
\end{array}\right)=\mathbb{T}_{L_{s}+1}\left(k\right)\cdot\mathbb{T}_{L_{s}}\left(k\right)\cdot\\
\cdots\cdot\mathbb{T}_{1}\left(k\right)\cdot\mathbb{T}_{0}\left(k\right)\cdot\left(\begin{array}{c}
\psi_{0}\\
\psi_{-1}
\end{array}\right).\label{TransferMatrixOneStep-1}
\end{multline}

In the same way, we may write the boundary conditions of Eqs.\,(\ref{BoundaryConditions}),
as the following matrix relations:

\vspace{-0.7cm}

\begin{equation}
\begin{array}{c}
\left(\begin{array}{c}
\psi_{0}\\
\psi_{-1}
\end{array}\right)=\underset{\mathbb{B}_{L}\left(k\right)}{\underbrace{\left(\begin{array}{cc}
e^{-ik} & e^{ik}\\
e^{-2ik} & e^{2ik}
\end{array}\right)}}\cdot\left(\begin{array}{c}
\Psi_{+}^{L}\\
\Psi_{-}^{L}
\end{array}\right),\\
\\
\end{array}\label{BoundaryLeft}
\end{equation}

\noindent and

\vspace{-0.8cm}

\begin{equation}
\left(\begin{array}{c}
\psi_{L_{s}+2}\\
\psi_{L_{s}+1}
\end{array}\right)=\left(\begin{array}{cc}
e^{2ik} & e^{-2ik}\\
e^{ik} & e^{-ik}
\end{array}\right)\cdot\left(\begin{array}{c}
\Psi_{+}^{R}\\
\Psi_{-}^{R}
\end{array}\right),
\end{equation}
which can be inverted as

\vspace{-0.6cm}
\begin{equation}
\left(\begin{array}{c}
\Psi_{+}^{R}\\
\Psi_{-}^{R}
\end{array}\right)=\mathbb{B}_{R}\left(k\right)\cdot\left(\begin{array}{c}
\psi_{L_{s}+2}\\
\psi_{L_{s}+1}
\end{array}\right)\label{BoundaryRight}
\end{equation}

Using Eqs.\,(\ref{BoundaryLeft}) and (\ref{BoundaryRight}) into
Eq.\,(\ref{TransferMatrixOneStep-1}), we get to the following final
result:

\vspace{-0.6cm}

\begin{multline}
\left(\begin{array}{c}
\Psi_{+}^{R}\\
\Psi_{-}^{R}
\end{array}\right)=\mathbb{B}_{R}\left(k\right)\cdot\mathbb{T}_{L_{s}+1}\left(k\right)\cdot\mathbb{T}_{L_{s}}\left(k\right)\cdot\\
\cdots\cdot\mathbb{T}_{1}\left(k\right)\cdot\mathbb{T}_{0}\left(k\right)\cdot\mathbb{B}_{L}\left(k\right)\cdot\left(\begin{array}{c}
\Psi_{+}^{L}\\
\Psi_{-}^{L}
\end{array}\right),
\end{multline}
and, by definition, the transfer matrix of the whole sample is written
as:

\vspace{-0.8cm}

\begin{multline}
\mathcal{M}\left(k\right)=\mathbb{B}_{R}\left(k\right)\cdot\mathbb{T}_{L_{s}+1}\left(k\right)\cdot\mathbb{T}_{L_{s}}\left(k\right)\cdot\\
\cdots\cdot\mathbb{T}_{1}\left(k\right)\cdot\mathbb{T}_{0}\left(k\right)\cdot\mathbb{B}_{L}\left(k\right).
\end{multline}
This last equation was the one we implemented to calculate $\mathcal{M}\left(k\right)$
for any given disordered sample.

$ $

\section{\label{appx:SelectionRule}Emergence of selection rule}

In this appendix, we prove the effective selection rule of Eq.~(\ref{eq:Effective_SelectionRule}).
In order to do so, we will analyze the factor $\sin\left[\left(k-q\right)L_{l}\right]$,
when $q,k$ belong to the same or different classes. More precisely,
will calculate its absolute value, which can be written as
\begin{widetext}
\vspace{-0.8cm}

\begin{multline}
\abs{\sin\left[\left(k-q\right)L_{l}\right]}=\sqrt{\cfrac{1-\cos\left[2L_{l}\left(q-k\right)\right]}{2}}\\
=\frac{1}{\sqrt{2}}\left\{ 1-\cos\left[2\left(L_{l}+1\right)\left(q-k\right)+\phi\left(q\right)-\phi\left(k\right)\right]\cos\left[\phi\left(q\right)-\phi\left(k\right)-2\left(q-k\right)\right]\right.\\
-\left.\sin\left[2\left(L_{l}+1\right)\left(q-k\right)+\phi\left(q\right)-\phi\left(k\right)\right]\sin\left[\phi\left(q\right)-\phi\left(k\right)-2\left(q-k\right)\right]\right\} ^{\frac{1}{2}},
\end{multline}
where we summed and subtracted $\phi\left(q\right)-\phi\left(k\right)$
in the argument of the cosine and, then, decomposed it using the rule
for the cosine of a sum of angles. The main advantage of this form
is that the continuous function $\phi\left(k\right)$ depends solely
in the properties of the central sample and the effect of increasing
the leads is to populate more densely their domains with allowed values
of $k$. This, together with the fact that we are only interested
in what happens near $k_{\text{F}}$, allows us to expand it as Taylor
series on $\delta q=q-k_{\text{F}}$ and $\delta k=k_{\text{F}}-k$:

\begin{equation}
\phi\left(q\right)-\phi\left(k\right)=\left.\frac{d}{dk}\phi\right|_{k_{\text{F}}}\left(\delta q+\delta k\right)+\cdots\simeq\left.\frac{d}{dk}\phi\right|_{k_{\text{F}}}\left(q-k\right),
\end{equation}
and, consequently,

\vspace{-0.5cm}

\begin{align}
\abs{\sin\left[\left(q-k\right)L_{l}\right]}\simeq & \sqrt{\cfrac{1-\cos\left[2L_{l}\left(q-k\right)+\phi\left(q\right)-\phi\left(k\right)\right]}{2}},\label{abssin2}
\end{align}
where the corrections are of order $q-k$ and disappear in the limits
$L_{l}\to\infty$ and $\hbar t^{-1}\to0$. At this point, all we must
do is to decompose the cosine term in Eq.\,(\ref{abssin2}) using
the usual rules for the sum of angles and then resort to the quantization
condition of Eq.\,(\ref{eq:QuantizationCondition}) to realize that

\vspace{-0.5cm}

\begin{multline}
\cos\left[2\left(L_{l}+1\right)\left(q-k\right)+\phi\left(q\right)-\phi\left(k\right)\right]\\
=\mp\sqrt{\left[1-\abs{r\left(q\right)}^{2}\sin^{2}\left(\theta\left(q\right)-\phi\left(q\right)\right)\right]\left[1-\abs{r\left(k\right)}^{2}\sin^{2}\left(\theta\left(k\right)-\phi\left(k\right)\right)\right]}\\
+\abs{r\left(q\right)}\abs{r\left(k\right)}\sin\left[\theta\left(q\right)-\phi\left(q\right)\right]\sin\left[\theta\left(k\right)-\phi\left(k\right)\right],\label{CosDecomposition}
\end{multline}
where the $+\left(-\right)$ sign stands for the case when $q$ and
$k$ are in the same class (different classes) of states.

Finally, one can evoke the same argument as before to Taylor expand
all the sample-specific functions appear in Eq. (\ref{CosDecomposition})
{[}to be clear, $r\left(x\right)$,$\theta\left(x\right)$,$\phi\left(x\right)${]}
around $k_{\text{F}}$, but noting that $k<k_{\text{F}}<q$ by definition.
Up to corrections irrelevant correction in the same limits, this gives
rise to Eq. (\ref{eq:Effective_SelectionRule}) of the main text after
expanding the $\sin$ functions in powers of $q-k$.

$ $
\end{widetext}

\section{\label{appx:CalculationJDoCS}Calculation of the joint density of
contributing states}

In this appendix, we will proceed to calculate the joint density of
contributing states (JDoCS), for both positive and negative $\Delta\varepsilon$.
For positive energy differences, $\Delta\varepsilon>0$, the JDoCS
is defined, from Eq.\,(\ref{eq:JDoCS}), as

\vspace{-0.5cm}

\begin{equation}
\varrho\left(\epsilon_{\text{F}},\Delta\varepsilon\right)=\frac{1}{4L_{l}^{2}}\sideset{}{^{'}}\sum_{\overset{k,q}{(\varepsilon_{q}<\varepsilon_{\text{F}}\leq\varepsilon_{k})}}\delta\left(\Delta\varepsilon-\Delta\varepsilon_{k,q}\right),\label{eq:JDoCS-1}
\end{equation}
which may be written in terms of the usual density of states for each
class, $\sigma=\pm$, i.e.,

\vspace{-0.6cm}
\begin{align}
\rho^{\sigma}\left(\varepsilon\right) & =\frac{1}{L_{l}}\sum_{k^{\sigma}}\delta\left(\varepsilon-\varepsilon_{k^{\sigma}}\right),\label{DoSDefinition}
\end{align}
yielding the expression,

\vspace{-0.5cm}
\begin{multline}
\varrho\left(\varepsilon_{\text{F}},\Delta\varepsilon\right)=\frac{1}{4}\int_{\varepsilon_{F}}^{2}d\epsilon_{2}\int_{-2}^{\varepsilon_{F}}d\epsilon_{1}\lim_{L_{l}\to\infty}\left\{ \rho_{L_{l}}^{+}\left(\epsilon_{1}\right)\rho_{L_{l}}^{-}\left(\epsilon_{2}\right)\right.\\
\left.+\rho_{L_{l}}^{-}\left(\epsilon_{1}\right)\rho_{L_{l}}^{+}\left(\epsilon_{2}\right)\right\} \delta\left(\Delta\varepsilon-\epsilon_{2}+\epsilon_{1}\right),\label{eq:DPED}
\end{multline}
in the limit of semi-infinite leads.

To progress beyond Eq.\,(\ref{eq:DPED}) in a general fashion, one
starts by recognizing that, since $\rho^{\pm}\left(\varepsilon\right)$
is an intensive quantity. So these must be dominated by the states
on the (clean) leads, as $L_{l}\to\infty$. Since we know that, for
a clean system, the states of different parities are alternated in
$k$-space, with a regular separation given by $\pi/L_{l}$, one concludes
that

\vspace{-0.7cm}

\begin{equation}
\lim_{L_{l}\to\infty}\rho_{L_{l}}^{\pm}\left(\varepsilon\right)=\rho\left(\varepsilon\right)=\begin{cases}
\frac{1}{\pi\sqrt{4w^{2}-\varepsilon^{2}}} & \text{if}\abs{\varepsilon}\leq2w\\
0 & \text{if}\abs{\varepsilon}>2w
\end{cases},\label{eq:DOS1D}
\end{equation}

\noindent where $\rho\left(\varepsilon\right)$ is the full DoS of
a clean infinite chain.

\begin{figure}[H]
\begin{centering}
\includegraphics[scale=0.35]{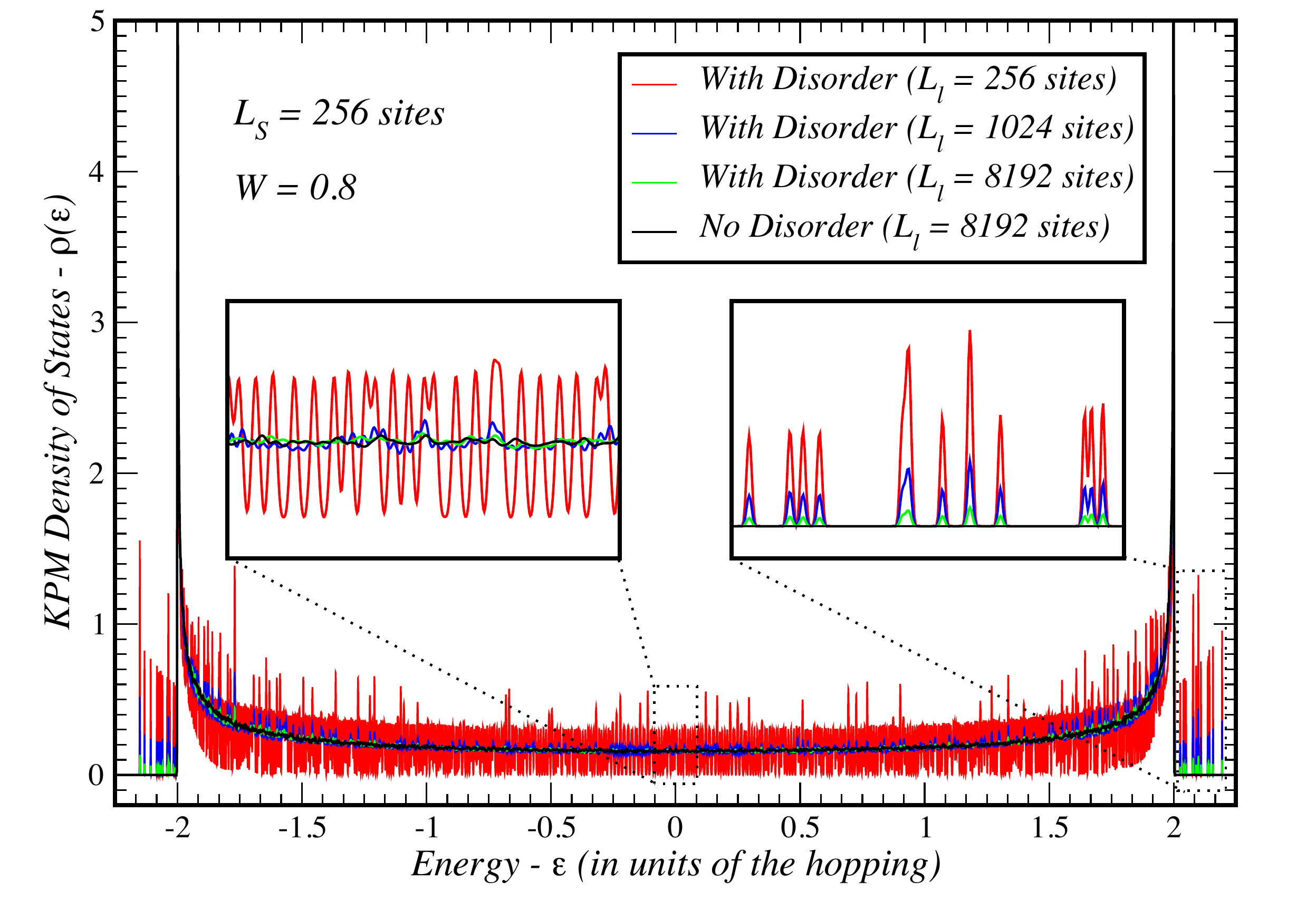}
\par\end{centering}
\caption{\label{DOSCovergence}Plots of the DoS calculated using the KPM for
a system with leads of different sizes and a central sample without
(black curve) and with disorder (colored curves). The number of Chebyshev
moments used is $M=4096$ for all the cases. The insets are zooms
made to the regions indicated by the black boxes in the main graph,
where one can clearly see the spectral weight of the states in the
sample being out-weighted by the states coming from the finite clean
leads.}
\end{figure}

\begin{figure}[H]
\begin{centering}
\includegraphics[scale=0.345]{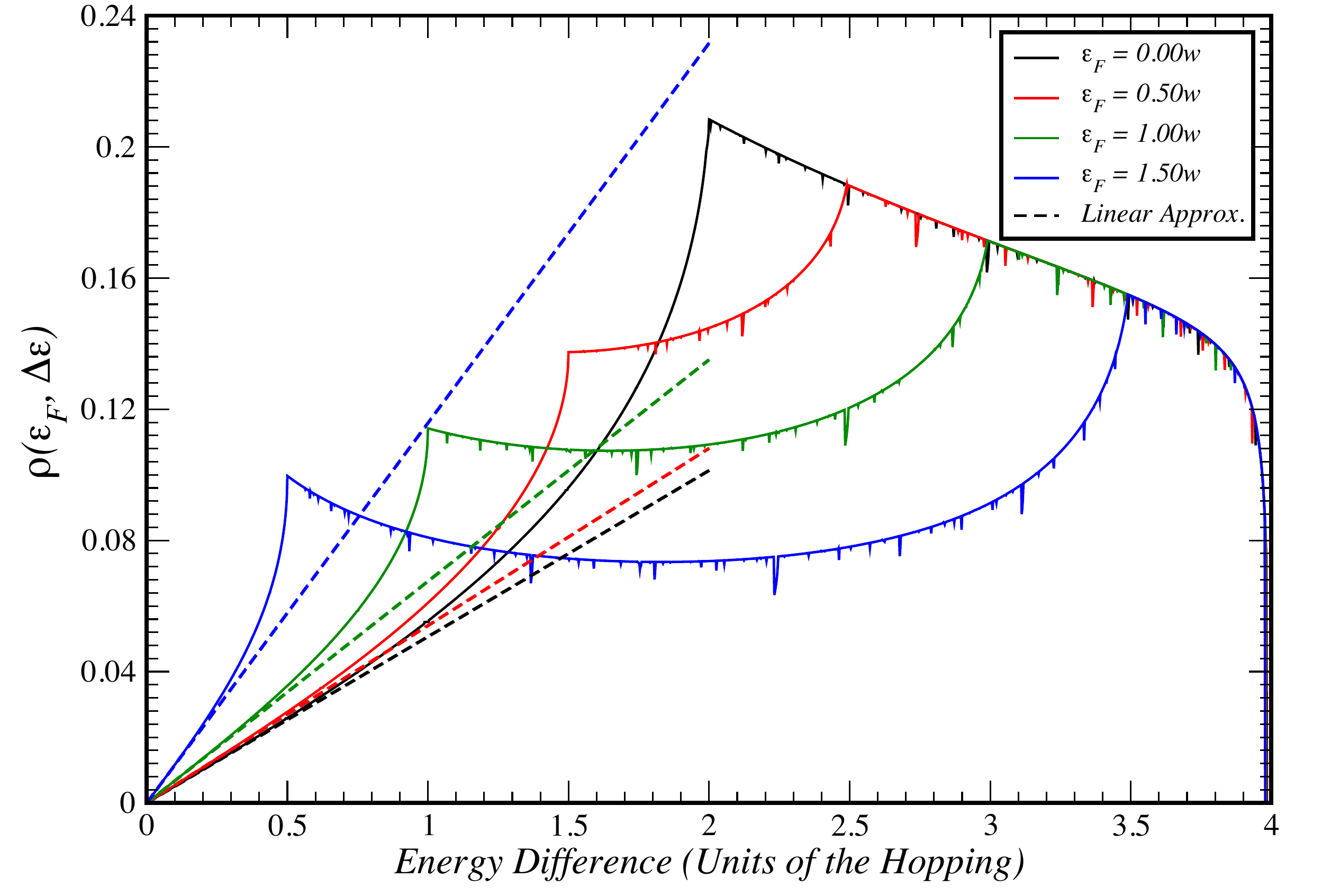}
\par\end{centering}
\caption{\label{DPED}Plots of the JDoCS from the numerical integration of
Eqs.\,(\ref{eq:DPED-1-1}) for different values of the Fermi energy
and positive values of $\Delta\varepsilon$. The dashed straight lines
are plots of the linear approximations near $\varepsilon_{\text{F}}$,
as calculated in Eq.\,(\ref{LinearizedDPEP-1}). (color online)}
\end{figure}

In what follows, we will always assume that the expression of Eq.\,(\ref{eq:DOS1D})
may be used to calculate de JDoCS in the limit of very large $L_{l}$.
This intuition is confirmed by the plots of the DoS in Fig.\,\ref{DOSCovergence},
which were obtained numerically, for a randomly selected disordered
sample, using the well-known\emph{ kernel polynomial method} with
a Jackson kernel and a fixed number of polynomials, $M=4096$, enough
to resolve the individual energy levels in the smaller case considered
(see Weiße\emph{ et al}. \citep{weise_kernel_2006} for more details
on the method). Consequently, one has the following expression for
the JDoCS:

\noindent \vspace{-0.6cm}{\small{}
\begin{align}
\varrho\left(\varepsilon_{\text{F}},\Delta\varepsilon\right) & =\int_{\varepsilon_{\text{F}}}^{2}d\varepsilon_{2}\frac{\Theta\left(\abs{\Delta\varepsilon}+\varepsilon_{\text{F}}-\varepsilon_{2}\right)}{2\pi^{2}\sqrt{\left(4w^{2}-\varepsilon_{2}^{2}\right)\left(4w^{2}-\left(\varepsilon_{2}+\Delta\varepsilon\right)^{2}\right)}}\nonumber \\
 & =\int_{\varepsilon_{\text{F}}}^{\varepsilon_{\text{F}}+\abs{\Delta\varepsilon}}\qquad\qquad\thinspace\thinspace\thinspace\qquad\qquad\qquad\qquad\qquad\mathllap{d\varepsilon_{2}\frac{\Theta\left(\abs{\Delta\varepsilon}+\varepsilon_{\text{F}}-\varepsilon_{2}\right)}{2\pi^{2}\sqrt{\left(4w^{2}-\varepsilon_{2}^{2}\right)\left(4w^{2}-\left(\varepsilon_{2}+\Delta\varepsilon\right)^{2}\right)}}}\label{eq:DPED-1-1}
\end{align}
}where $\Theta\left(x\right)$ is the Heaviside function and $\Delta\varepsilon\geq0$.
The integral in Eq.\,(\ref{eq:DPED-1-1}) can be done numerically
and the curves are shown in Fig.\,\ref{DPED} for different values
of the Fermi energy $\epsilon_{\text{F}}$. Nevertheless, we are only
interested in the shape of $\varrho\left(\varepsilon,\Delta\varepsilon\right)$
when $\Delta\varepsilon\approx0$. For that, we may expand Eq.\,(\ref{eq:DPED-1-1})
in powers of this quantity, yielding

\noindent \vspace{-0.7cm}

\begin{equation}
\varrho\left(\varepsilon_{\text{F}},\Delta\varepsilon>0\right)=\frac{\Delta\varepsilon}{2\pi^{2}\left(4w^{2}-\varepsilon_{\text{F}}^{2}\right)}+\mathcal{O}\left[\Delta\varepsilon^{2}\right].\label{LinearizedDPEP-1}
\end{equation}

Finally, we can generalize Eq.\,(\ref{LinearizedDPEP-1}) to $\Delta\varepsilon<0$,
which is trivial since, by definition {[}Eq.\,(\ref{eq:JDoCS}){]},
we have $\varrho\left(\Delta\varepsilon\right)=\varrho\left(-\Delta\varepsilon\right)$.
Hence, our final expression is simply,

\noindent \vspace{-0.7cm}

\begin{equation}
\varrho\left(\epsilon_{\text{F}},\Delta\varepsilon\right)=\frac{\abs{\Delta\varepsilon}}{2\pi^{2}\left(4w^{2}-\varepsilon_{\text{F}}^{2}\right)}+\mathcal{O}\left[\Delta\epsilon^{2}\right],\label{LinearizedDPEP-1-1}
\end{equation}
which is the one we use in the main text {[}see Eq.~(\ref{LinearizedDPEP}){]}.

\noindent 

\noindent 

$ $

$ $

$ $

$ $

$ $

$ $

\bibliographystyle{apsrev4-1}
\bibliography{References_NEW}

\end{document}